\documentclass[aps,prb,reprint,floatfix,superscriptaddress,amssymb,amsmath]{revtex4-2}
\usepackage{color}
\usepackage{graphicx}
\usepackage{esint}
\usepackage{times}

\usepackage[hypertexnames=false,naturalnames=true]{hyperref}
\hypersetup{colorlinks=True}
\usepackage[all]{hypcap}

\newcommand{\beq}{\begin{equation}}
\newcommand{\eeq}{\end{equation}}
\newcommand{\bea}{\begin{eqnarray}}
\newcommand{\eea}{\end{eqnarray}}

\newcommand{\phdag}{{\phantom{\dagger}}}
\newcommand{\psiLdag}{\psi^\dagger_L}
\newcommand{\psiL}{\psi^\phdag_L}
\newcommand{\psiRdag}{\psi^\dagger_R}
\newcommand{\psiR}{\psi^\phdag_R}
\newcommand{\HT}{H_\text{\,T}}
\newcommand{\Vg}{V_\text{gate}}

\newcommand{\sdfrac}[2]{\mbox{\small$\displaystyle\frac{#1}{#2}$}}

\begin{document}

\title{Conductance of a Dissipative Quantum Dot:\\Nonequilibrium Crossover Near a Non-Fermi-Liquid Quantum Critical Point}

\affiliation{Department of Physics, Duke University, Durham, North Carolina 27708-0305, U.S.A.} 
\affiliation{Centro de Ci\'{i}ncias Naturais e Humanas, Universidade Federal do ABC, Santo Andr\`{e}, SP 09210-580, Brazil}
\affiliation{Institute for Quantum Materials and Technologies, Karlsruhe Institute of Technology, 76021 Karlsruhe, Germany}

\author{Gu Zhang}
\email{gu.zhang@kit.edu}
\affiliation{Department of Physics, Duke University, Durham, North Carolina 27708-0305, U.S.A.} 
\affiliation{Institute for Quantum Materials and Technologies, Karlsruhe Institute of Technology, 76021 Karlsruhe, Germany}

\author{E. Novais}
\email{eduardo.novais@ufabc.edu.br}
\affiliation{Centro de Ci\'{i}ncias Naturais e Humanas, Universidade Federal do ABC, Santo Andr\`{e}, SP 09210-580, Brazil}

\author{Harold U. Baranger}
\email{baranger@phy.duke.edu}
\affiliation{Department of Physics, Duke University, Durham, North Carolina 27708-0305, U.S.A.}

\begin{abstract}
We find the nonlinear conductance of a dissipative resonant level in the nonequilibrium steady state near its quantum critical point. The system consists of a spin-polarized quantum dot connected to two resistive leads that provide ohmic dissipation. We focus on the crossover from the strong-coupling, non-Fermi-liquid regime to the weak-coupling, Fermi-liquid ground state, a crossover driven by the instability of the quantum critical point to hybridization asymmetry or detuning of the level in the dot. We show that the crossover properties are given by tunneling through an effective single barrier described by the boundary sine-Gordon model. The nonlinear conductance is then obtained from thermodynamic Bethe ansatz results in the literature, which were developed to treat tunneling in a Luttinger liquid. The current-voltage characteristics are thus found for any value of the resistance of the leads. For the special case of lead resistance equal to the quantum resistance, we find mappings onto, first, the two-channel Kondo model and, second, an effectively noninteracting model from which the nonlinear conductance is found analytically. A key feature of the general crossover function is that the nonequilibrium crossover driven by applied bias is different from the crossover driven by temperature---we find that the nonequilibrium crossover is substantially sharper. Finally, we compare to experimental results for both the bias and temperature crossovers: the agreement is excellent.
\end{abstract}

\date{30 September 2021; accepted version of manuscript published as \href{https://doi.org/10.1103/PhysRevB.104.165423}{Phys.\,Rev.\,B \textbf{104}, 165423 (2021)}}
\maketitle

\section{Introduction and Summary}
\label{sec:intro}

This work addresses nonequilibrium properties of correlated electrons in a nanofabricated system, one in which dissipation causes the correlations, making it an open quantum system with a non-Markovian environment. Correlated electronic systems exhibit a wide variety of 
poorly understood phenomena and remain a major challenge in contemporary physics \cite{SCEScollection17, ProcSCES2019, FuldeBook}.  Sudden qualitative changes in the ground state as a function of tuning parameters, called quantum phase transitions, often occur \cite{CarrBook,SachdevBook,VojtaPhilMag06}.  
At the transition---the quantum critical point (QCP)---novel correlated states are observed: these are typically referred to as ``non-Fermi liquids'' in that they do not follow the Fermi-liquid paradigm for normal metallic states.  
A common feature of most systems is that the correlations are produced by operators (interactions) that cause frustration 
and act locally.  A thorough understanding of systems in which correlating operators act in a single local region is, then, an essential foundation---such a system is known as a quantum impurity problem.  The ground states and equilibrium properties of many quantum impurity problems have been studied, including ones with quantum phase transitions and associated QCP \cite{VojtaPhilMag06,Vojta_critquasiEPJST15}. 
Perhaps the best known are the Kondo problems (original, multi-channel, and multi-impurity); indeed, the Kondo model has inspired a myriad of different physical problems and fundamental advances in theoretical physics \cite{Vojta_critquasiEPJST15, HewsonBook, CoxZawadowskiAdvanPhys98, ColemanBook}.  While the original isotropic antiferromagnetic Kondo problem and related Anderson model do not have a quantum phase transition (the ground state is always a Fermi liquid), the two-channel Kondo (2CK) model, for instance, does have a transition and associated QCP caused by quantum frustration. 
In nanoscale systems, such correlated electron physics, as well as nonequilibrium phenomena and dissipation, all arise naturally and under conditions that can be controlled.  

The confinement of electrons to nanoscale objects enhances electron-electron interaction effects.  Indeed, many confined nanoscale structures are inherently strongly correlated many-body problems and exhibit counterintuitive properties. 
The astonishing development in the construction and characterization of low-dimensional quantum systems over the last few decades has lead to exquisite control over their parameters. 
As a result, the use of strongly correlated nanoscale structures in nanoelectronics and spintronics is being pursued, which requires a fundamental understanding of the interplay between transport properties and interactions. 
Furthermore, the degree of control and quality of fabrication in nanosystems gives rise to the possibility of engineered interactions, making quantum simulation possible (see, e.g., 
\cite{DesjardinsKontosNature17, AnthorePierrePRX18, VandersypenNagFerroNature20, LevySerpentineSciAdv20, Frolov3AndreevX21, Veldhorst2DarrayAPL21}).   
As there are typically only a single or a few active regions in nanoscale systems---which is where interactions occur---they are natural quantum impurity problems. 

Nonequilibrium phenomena readily occur in nanoscale systems: a modest voltage applied to a nanoscale device results in a large field.  Indeed, measurements of nonequilibrium current-voltage relations ($I$-$V$ curves) are ubiquitous \cite{IhnBook}.  
Rapid time-dependent phenomena, such as a quantum quench protocol in which system parameters are suddenly changed, are harder to study in nanoscale systems, but the recent hybrid microwave-nanoscale systems raise possibilities in this direction.
More generally, nonequilibrium many-body phenomena have attracted increasing attention over the last decade but, compared to equilibrium or ground state phenomena, are much less well understood \cite{CarrBook, PolkovnikovNoneqClosedRMP11, NoneqDMFT-RMP14, EisertNoneqreviewNatP15, LeHurDrivenQImpReviewCRP18, RylandsNoneqIntegrable20}. Here, we focus on the nonequilibrium nonlinear $I$-$V$ curve of our nanoscale system. 

Dissipation also occurs very naturally in nanoscale systems as typically they are surrounded by many other degrees of freedom, referred to as a bath or environment. Dissipation and decoherence are usually to be avoided or at least minimized.
This is, however, not always the case, for dissipation is known to trigger boundary quantum phase transitions through the reduction of possible obstructions \cite{Mebrahtu12,Zhangdiss2IK17}.
Our system is indeed one in which
dissipation is a key part of the quantum impurity: it is an open many-body quantum system, a topic of great current interest \cite{BreuerPetruccioneBook, deVegaAlonsoRMP17}, with an environment that is inherently non-Markovian.

\begin{figure}[t]
\includegraphics[width=0.40\textwidth]{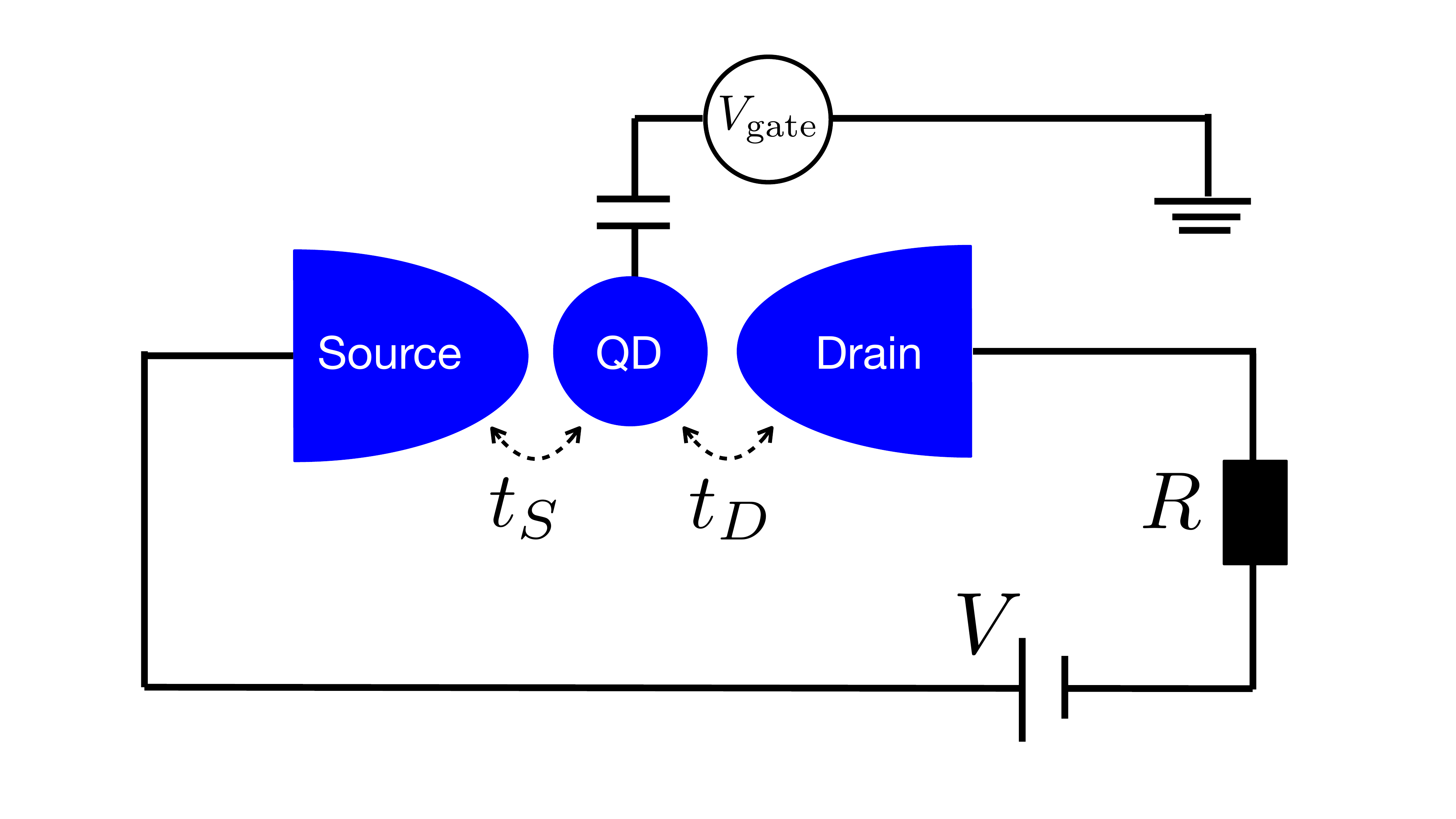} 
\caption{The setup of the dissipative resonant level (DRL) model: two leads, source (S) and drain (D), couple to a quantum dot (QD) with hopping strengths $t_S$ and $t_D$, respectively. The dissipative environment is represented by an ohmic impedance $R$. Bias $V$ is applied between the two leads, and the quantum dot energy level is tuned by the backgate voltage $V_{\text{gate}}$. 
}
\label{fig:model} 
\end{figure}

The system we study is a quantum dot connected to resistive leads through tunable tunnel barriers as sketched in Fig.\,\ref{fig:model}. 
The resistance of the leads, denoted $R$, creates an ohmic dissipative environment \cite{Bomze09,Mebrahtu12}, and the quantum dot serves as the quantum impurity.  The most remarkable feature of this system is the existence of a frustration-induced non-Fermi-liquid QCP: it occurs when both (i) a level in the dot is resonant with the leads and (ii) the dot is symmetrically coupled to them. 
Both of these properties can be fine-tuned using gate voltages, enabling experimental access to the QCP.  At the QCP, which is of the 2CK type \cite{Mebrahtu12}, the conductance through the dot becomes perfect ($e^2/h$ at zero temperature) while otherwise it tends to zero. We previously presented several scaling relations, including non-Fermi-liquid scaling  
measured as a function of temperature, $T$, for negligible bias (i.e.\ in the equilibrium regime) \cite{Mebrahtu12,Mebrahtu13}, as well as nonequilibrium scaling at the QCP \cite{ZhangNoneqPRR21}. 
The presence of an \emph{interacting} QCP suggests viewing the problem through the lens of the renormalization group (RG)  \cite{FisherRGRMP98}: below we discuss scaling, relevant/irrelevant operators, and RG flows. 

Related systems have been studied with similar methods, and we make extensive use of previous literature. First, there is a close connection between the system studied here and resonant tunneling (through a double barrier) in a Luttinger liquid (LL), i.e.\ an interacting one-dimensional (1D) system.  While there is nothing explicitly 1D in our system, quantum impurity problems are all effectively 1D because of the local 
nature of the interaction operators. Indeed, tunneling in a resistive environment has been viewed as a quantum simulation of tunneling in a LL with repulsive interactions 
\cite{MatveevGlazman93,FlensbergPRB93,SassettiWeissEPL94,SafiSaleurPRL04,LeHurLiPRB05, BordaPRB05,*BordaZarandX06, FlorensPRB07,  Mebrahtu12, Mebrahtu13, JezouinPierre13, AnthorePierrePRX18, LeHurDrivenQImpReviewCRP18}.
In linear response, resonant peaks of perfect conductance in a LL have been extensively studied theoretically   \cite{KaneFisherPRB92,*KaneFisherPRB92a,EggertAffleck92,Furusaki93,*Furusaki98,YiKanePRB98,*YiPRB02,NazarovGlazman03,polyakov03,KomnikGogolinPRL03,Meden2005,GoldsteinPRL10a,HuKaneX16}. There is a 2CK-like QCP separating weak-tunneling regimes dominated by one of the tunneling barriers, either source or drain \cite{EggertAffleck92,YiKanePRB98,KomnikGogolinPRL03}. 

Nonlinear $I$-$V$ characteristics have been used to study the interplay between nonequilibrium and many-body effects in a variety of nanosystems \cite{AlhassidRMP00, ChangLLrevRMP03, Micolich0.7Rev11, LevyYeyatiSqdotS-rev11, DavidGG_CarrBook, AnthoreUniversalityRevEPJST20}. 
Experimental systems studied include the Kondo effect in quantum dots, tunneling into edge channels, and dissipative tunneling.  Theoretically, in the scaling regime in which $I\propto V^\alpha$, the exponent $\alpha$ has been frequently deduced from the scaling dimension of the leading operators at the QCP (see for example \cite{FisherGlazman97,AffleckLesHouches10}).  For resonant tunneling in a LL, for instance, the scaling power-law was obtained in this way in early work \cite{KaneFisherPRB92, EggertAffleck92}.  
A few full calculations beyond the scaling exponent exist in the literature. First, approximate numerical treatments have been employed \cite{HettlerPRL94,*HettlerPRB98,vonDelftAnPhys99,BuxboimPRB03,KirchnerSiPRL09, LeeChungPRB13, LandauCornfeldSelaPRL18}, though not of the model we study. Second, analytical $I$-$V$ curves have been obtained for the crossover from a QCP to a Fermi liquid state for the two-impurity, two-channel, and topological Kondo models \cite{SelaExactTransPRL09,*SelaNoneqQdotsPRB09, MitchellSelaPRL16, BeriPRL17}. 
These latter are particularly relevant here and are discussed further below.

\begin{figure}[b]
\includegraphics[width=0.48\textwidth]{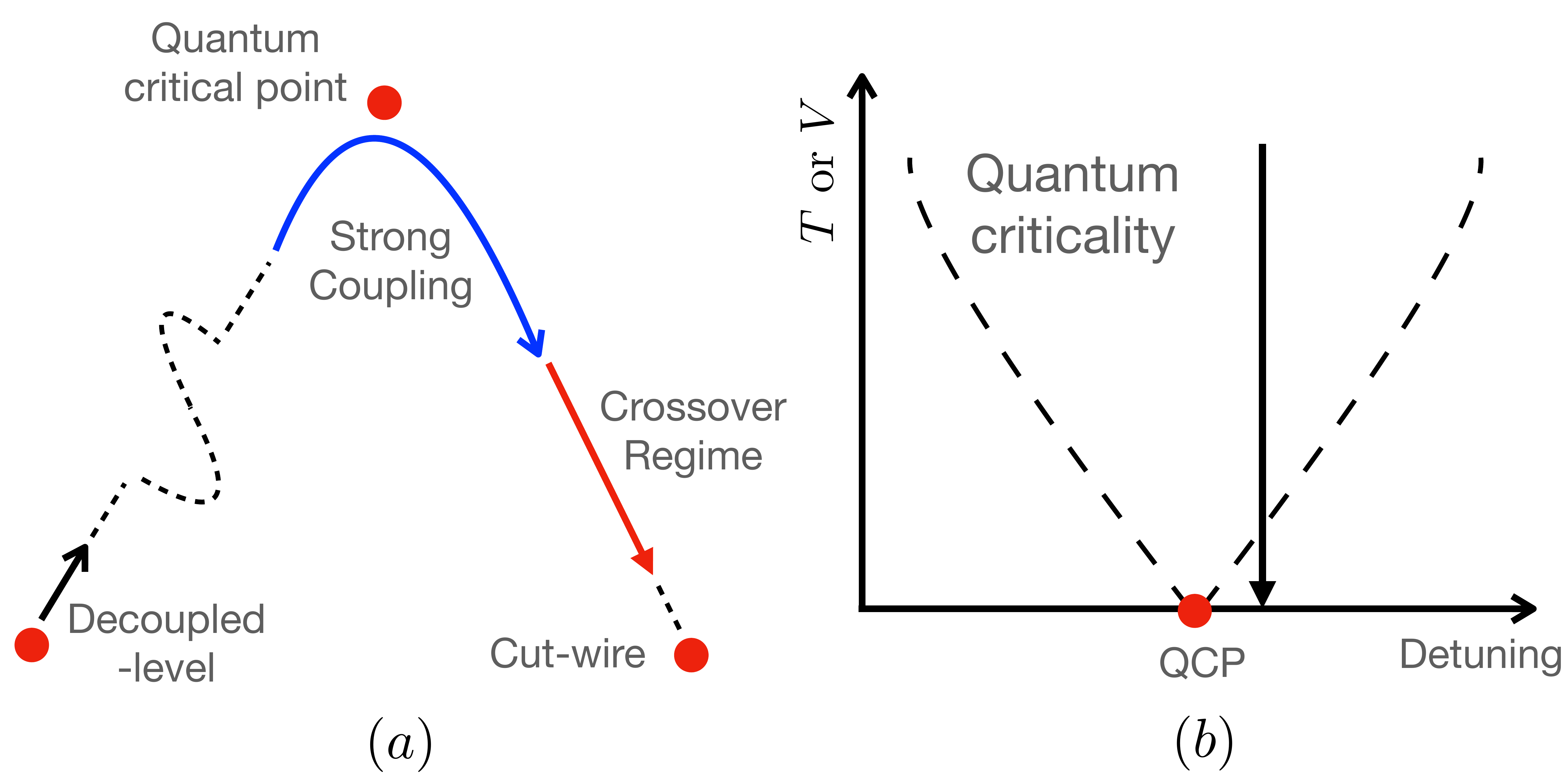} 
\caption{(a) Schematic RG flow of the system with three fixed points marked in red. Coupling strength near the decoupled-level fixed point (with Hamiltonian given in App.\,\ref{sec:Hweakcoup}) increases with the RG flow, driving the system towards the strong coupling regime (indicated by the blue arrow) described by Eq.\,\eqref{eq:Hboson_full}. Although the system being close to the quantum critical point, in this regime the system asymmetry or detuning becomes important. At lower energy scales, the system enters the crossover regime (the red arrow), and flows towards the cut-wire fixed point at which the conductance vanishes. All three fixed points can be distinguished by their impurity entropy. (b) The corresponding flow in the phase diagram. The system begins to enter the crossover regime near the quantum critical boundary.
}
\label{fig:flow} 
\end{figure}

The main result presented here is the calculation of the nonequilibrium nonlinear $I$-$V$ curve of this dissipative resonant level (DRL) model when the system is tuned very near to but not exactly at the QCP. In this case, the renormalization flow is first toward the full-conductance quantum critical point (strong-coupling) but then veers away and ultimately approaches the single barrier, low-conductance cut-wire fixed point (weak-coupling), as illustrated in Fig.\,\ref{fig:flow}. That is, we focus on the \textit{crossover} from the strong-coupling to weak-coupling behavior.

Crossover denotes the regime connecting two physically distinct fixed points \cite{CardyCFT84,VojtaPhilMag06}. The properties of the system first reflect being in the neighborhood of one fixed point (unstable) and then at lower energy scales---reached by decreasing the temperature or bias---come to reflect a second (stable) fixed point. Examples of crossover behavior in nanoscale physics include the connection between the strong and weak coupling fixed points in a Luttinger liquid \cite{LoCrossoverPRB19}, the non-Fermi-liquid and Fermi-liquid fixed points in a 2CK model \cite{FabrizioCrossoverPRL95}, and the free spin and screened spin fixed points in a spin boson model.
In the crossover regime, fluctuations in the system, both quantum and thermal, strongly influence the many-body interactions. 
An interesting aspect of crossover phenomena is that they are less universal: different ways of introducing a low energy scale, such as temperature vs bias, can yield different results. For the simpler problem of tunneling through a single barrier in a LL, different crossover behavior as a function of $T$ or $V$ has been connected \cite{AristovWoelfle09} to different third-order interaction contributions to the RG flow. For a resonant system, however, treatment of this topic is still absent. 

We find the $I$-$V$ curve by showing how our system in the crossover regime connects to the problem of tunneling through a single barrier. Indeed, from the physical two barrier system, an effective single barrier problem emerges at low energy in which left-moving particles tunnel into right-moving ones and vice versa. We then use results for tunneling through a single barrier in the presence of dissipation or interactions \cite{GiamarchiBook,GogolinBook,AnthoreUniversalityRevEPJST20}---in particular those of Fendley, et al.\,\cite{FendleySaleurWarnerNPB94,FendleyPRL95,FendleyPRB95,FendleyUnifiedFrameJSP96,FendleySaleur96}---to find the $I$-$V$ curve. This is plotted and compared to experimental measurements (see Fig.\,\ref{fig:sine_gordon}): the agreement is excellent. 

We emphasize that the strength of the dissipation is \emph{arbitrary}, as is necessary to compare to experiment (for which $R \!\approx\! 0.75 h/e^2 $). On the other hand, for a special value, $R\!=\!h/e^2 $, the effective single barrier problem is \textit{noninteracting}. In that case, in contrast, analytic expressions for the $I$-$V$ curve are obtained [Eq.\,(\ref{eq:toulouse_current})].
This special case is similar to the previous work done for the two-impurity, two-channel, and topological Kondo models \cite{SelaExactTransPRL09,*SelaNoneqQdotsPRB09, MitchellSelaPRL16, BeriPRL17} in which exact nonequilibrium crossover results were obtained for several interacting quantum impurity models by finding mappings to effectively noninteracting single-barrier problems. In our case, no such mapping is known, forcing us to treat the interacting problem. 

Very recently, exact results for the interacting single-barrier problem have been obtained for several special values of the dissipation \cite{boulatX2019}, using an approach quite different from that in Ref.\,\cite{FendleyPRB95} on which our work is based. The extensive comparison to experiment is impressive \cite{AnthorePierrePRX18}. 
However, in the interacting single barrier problem, the quantum phase transition occurs when the interaction is tuned to zero \cite{KaneFisherPRB92}, implying that there is no quantum critical state but rather a Fermi liquid. The crossover studied \cite{AnthorePierrePRX18} is therefore from a Fermi liquid to the low energy interacting state.  
In contrast, the system here has an interacting QCP: there is first RG flow toward a strong-coupling point and then away from it. This allows us to study the crossover from one interacting problem to another, which is the typical scenario for a quantum phase transition. 

A key feature of our result, $I(V,T)$, is that the dependence on the applied bias $V$ and temperature $T$ is different. At first sight, through the lens of RG, this seems counterintuitive: both $V$ and $T$ introduce a low energy cutoff on the RG flow, suggesting that the dependence may be similar. Indeed, in the asymptotic regimes, when parameters are such that one samples the behavior near only \emph{one} fixed point (either one), properties are controlled by the scaling dimension of the leading operator near that point and so the dependence on $V$ or $T$ is identical. 
However, there is no reason that the crossover function connecting these two asymptotic regimes should be the same \cite{KaneFisherPRB92, AristovWoelfle09, AnthorePierrePRX18}. We calculate both crossover functions and show, in good agreement with experiment, that in the nonequilibrium case (bias dependence at $T\!=\!0$) the transition from one asymptotic behavior to the other is substantially sharper. 

The organization of the paper is as follows. We start in Sec.\,\ref{sec:model-hamiltonian} by presenting the physical system, the initial weak link model, and the main features of the RG flow. 
Sec.\,\ref{sec:strong-coup} contains the key derivation of the effective description near the full-transmission (strong-coupling) quantum critical point. In Sec.\,\ref{sec:I-V}, the $I$-$V$ curves are calculated and compared to experiment. In addition, theoretical results for the special case $R\!=\!h/e^2$ are given. 
In Sec.\,\ref{sec:Discussion}, we discuss our physical understanding of the difference between the bias and temperature crossovers and the connection between the DRL and 2CK models. Conclusions are presented in Sec.\,\ref{sec:Conclusion}. 
Four appendices contain some details: the initial system Hamiltonian and bosonization for weak links (App.\,\ref{sec:Hweakcoup}), the mapping to the noninteracting problem for $R\!=\!h/e^2$ (App.\,\ref{sec:toulouse-point}), alternative plotting of the $R\!=\!h/e^2$ results (App.\,\ref{sec:other_toulouse}), and the explicit mapping to the 2CK model (App.\,\ref{sec:2CKmapping}).

\section{The System: Weak Link Model and Renormalization}
\label{sec:model-hamiltonian}

The system sketched in Fig.\,\ref{fig:model} has been implemented using a carbon nanotube quantum dot contacted by Cr/Au electrodes connected to Cr resistors to provide dissipation \cite{Mebrahtu12,Mebrahtu13,ZhangNoneqPRR21}. The energy level spacing in the dot is large, and a magnetic field is applied in order to spin-polarize the top level in the dot. The resistance of the leads creates an ohmic dissipative environment \cite{Bomze09}. 
In Fig.\,\ref{fig:model}, $t_{S}$ and $t_{D}$ denote the tunneling amplitudes between the dot and the two leads (source and drain). Experimentally, these two parameters are tunable through side gates. The occupation number of the dot can also be tuned through the effect of the gate voltage $\Vg$ on the energy of the top level in the dot. Finally, the source-drain bias driving the nonequilibrium steady state is denoted $V$. 

The strength of the dot-lead coupling ($t_S$ and $t_D$) is small experimentally: the system is in the \emph{weak link} limit, which is equivalent to a large tunneling barrier, and close to the (unstable) decoupled-level fixed point. The properties of the system in equilibrium are well established; we summarize them here and refer the reader to Appendix \ref{sec:Hweakcoup} for more detail and explanation. 

First, how does the environment enter the problem? The dissipative environment is characterized by an electromagnetic impedance which in our case is simply a resistance, $R$, because the spectrum is ohmic. By dividing by the quantum resistance $h/e^2$, we introduce the dimensionless resistance to characterize the environment, $r\equiv Re^2/h$. It is natural then that the environment is connected to the charge of the tunneling electron. When the electron tunnels, the charge on the capacitor formed by the tunnel junction shifts. This sudden change in charge excites the electromagnetic environment. The dot-lead tunneling amplitude is therefore accompanied by a charge shift operator, which makes the tunneling contribution to the Hamiltonian \emph{not} quadratic. Thus, effective interactions are generated among the electrons, mediated by the environment. 

We include only one level in the quantum dot and assume the leads are noninteracting, which renders the problem mathematically one dimensional \cite{HewsonBook,GogolinBook}. The explicit connection between tunneling in the presence of an environment and LL physics is made through the relation $g\!=\!1/(1+r)$ between the LL interaction parameter $g$ and our dimensionless resistance $r$ \cite{SafiSaleurPRL04, FlorensPRB07, Mebrahtu12, LiuRLdissipPRB14, Zheng1-GPRB14}. 
The interaction of the tunneling electrons with the environment causes a renormalization of the system parameters because of the effective electron-electron interaction. 
RG flow occurs as quantum fluctuations in the environment on lower and lower energy scales are taken into account \cite{FisherRGRMP98}. Experimentally, RG flows are related to changes in observed properties as a function of temperature (or applied bias in the nonequilibrium case) and have been measured in a few cases \cite{WeiTsuiPRB86, MurzinFQHEflowPRB05, IftikharPierreNat15,IftikharPierreScience18}. 

In the results reported here, we focus on the regime $r \alt 1$ in order to avoid situations in which the charge on the dot is localized \cite{LargeR}.
We assume that the junctions between the dot and the leads have the same capacitance, though the tunneling amplitude can differ (they are electrically symmetric but not necessarily quantum mechanically symmetric). This means that the environmental effect of a tunneling event from the source lead is the same as that to the drain.      

Previous work has shown that if the system is tuned such that it is both on resonance and symmetric in tunneling, $t_S = t_D$, then renormalization causes the effective tunneling between the dot and the leads to grow (barriers get smaller)  \cite{Mebrahtu12, Mebrahtu13, LiuRLdissipPRB14}. The system is described by an RG fixed point in which the dot is strongly hybridized with both leads and the source-drain transmission probability is unity, its maximum possible value. We refer to this as the full-transmission or ``strong-coupling'' fixed point. 

In contrast, if the system is either off resonance or asymmetric in dot-leads tunneling, then the modes of the environment suppress tunneling, as one would normally expect of a dissipative environment.
Renormalization (in most cases) causes an initially off-resonance level in the dot to become more off resonance, eventually becoming either completely filled or completely empty. 
Likewise an initial asymmetry $t_S \neq t_D$ grows such that the dot is incorporated into one of the leads while the link to the other lead is cut. The RG flow ends at a zero transmission (zero conductance) fixed point corresponding to a cut wire---two disconnected, semi-infinite 1D wires \cite{AsymmetricFP}. 

Here we suppose that the system is tuned very close to but not exactly at the resonant-symmetric condition. There is a small detuning from resonance and/or a small asymmetry in the dot-leads tunneling. The RG flow, shown schematically in Fig.\,\ref{fig:flow}, is then initially toward the full-transmission strong-coupling quantum critical point, where the renormalized level width is larger than the renormalized detuning. 
Once in the vicinity of the strong-coupling point, however, the small energies associated with the detuning come into play. There is a second stage of renormalization in which the flow is from the full-transmission to the cut-wire fixed point. The main goal of this paper is to describe transport through the quantum dot during this crossover from strong to weak coupling. 

\section{Effective Description Near Full Transmission (Strong Coupling)}
\label{sec:strong-coup}

To carry out our program, we need a model of the system near the full-transmission fixed point. Because the link between the dot and leads strengthens according to the initial RG equations that apply near the decoupled-level fixed point (see App.\,\ref{sec:Hweakcoup}), it is reasonable to assume that the initial weak-link model connects to the case of two small barriers in a 1D wire. Indeed, the constraint that the fixed point must be conformally invariant implies a certain universality and so freedom to choose a convenient starting point for the strong-coupling description \cite{AffleckMiniRev94, WongAffleck94, AffleckLesHouches10}. 

We therefore start with 1D noninteracting fermions described by right- and left-moving fields $\psi_R(x)$ and $\psi_L(x)$ with Fermi velocity set to one and add a static potential barrier consisting of two $\delta$-functions at $x \!=\! \pm\ell/2$. 
The most important effect of the potential on the fermions is to \emph{backscatter} them. Thus for the coupling between the dot and leads, we use
\begin{equation}\label{eq:HTfermion}
\HT = \sum_\pm A_\pm' \left[ \psiLdag(\pm\ell/2) \psiR(\pm\ell/2) + 
\text{h.c.}
\right] ,
\end{equation}
where $A_\pm'$ are assumed to be small. 

The possibility of detuning from resonance is realized through the voltage applied to the gate (see Fig.\,\ref{fig:model}), which couples to the density of fermions between the barriers: 
\begin{equation}\label{eq:Hgate1} 
H_\text{gate} = \varepsilon(\Vg) \!\int_{-\ell/2}^{+\ell/2} \!\!\!\!\!\! dx 
\left[ \psiLdag(x) \psiL(x)  + \psiRdag(x) \psiR(x) \right],
\end{equation}
where the coefficient $\varepsilon$ is controlled by the gate voltage with $\varepsilon \!=\! 0$ when the quantum dot is tuned to resonance, which we take to be $\Vg\!=\!0$. We have retained only smoothly varying terms in the density, assuming the potential produced by $\Vg$ is smoothed by the RG flow. 
A small rapidly varying $2k_F$ component in the density would cause $\Vg$-dependent backscattering that could be included in the barrier term \eqref{eq:HTfermion}.

$\varepsilon$ incorporates the energy level of the quantum dot
[denoted $\epsilon_\text{d}(\Vg )$ in App.\,\ref{sec:Hweakcoup}], 
but the exact relation between $\varepsilon$ and $\Vg$ is not known because it is the result of the renormalization process which we cannot follow exactly. 
The perturbative RG equations near the decoupled-level fixed point show that the effective detuning from resonance increases during renormalization (see App.\,\ref{sec:Hweakcoup}). Nevertheless, we assume that the initial detuning is so small that the RG flow approaches the strong-coupling point and thus $\varepsilon$ is small. Since experimentally the conductance approaches unity, the renormalized detuning remains smaller than the renormalized width of the level. 

We proceed via phenomenological bosonization in the standard way \cite{GiamarchiBook,KaneFisherPRB92,GogolinBook}, choosing the conventions of Ref.\,\cite{KaneFisherPRB92}:
\begin{equation}
\psi^{\dagger}_{L/R}(x,t) =  e^{\pm i k_F x} \frac{F}{\sqrt{2\pi a_0}} 
e^{i \sqrt{\pi} [\phi(x,t) \pm \theta(x,t)]}  
\label{eq:bosonization}
\end{equation}
where $\pm$ in the exponent corresponds to left- or right-moving particles, $L/R$.
$\phi(x)$ and $\theta(x)$ are conjugate bosonic operators that represent the fluctuations in the density and phase of the fermions. They obey the standard commutation relation $[\phi(x'),\partial_x\theta(x)]=i\pi\delta(x'-x)$. 
$a_{0}$ is a regularization scale for short distance or time, and $F$ is a Klein factor that can be simply carried along in the present problem, giving rise to no additional phase, and so we do not discuss it further. 
For the moment the fields are noninteracting---we add the environment as well as the applied bias below. 

The free fermionic Hamiltonian, $H_0$, becomes simply that for free bosons. 
The density of fermions appearing in  $H_\text{gate}$ is related to the bosonic fields by 
\begin{equation}
    \rho_{L/R}(x) = \frac{1}{2 \sqrt{\pi}} \big[ \pm \partial_x \phi(x) + \partial_x \theta(x) \big] 
+ \frac{k_F}{2\pi}.
\label{eq:density}
\end{equation}
Thus, in bosonic form \footnote{In writing these equations, 
we have chosen the bosonization convention near the strong-coupling fixed point (enhanced transmission implies weak barriers). In the following calculation, effects of $\HT$ are found to only leading order in $A$. In this way, the subtle concerns of bosonization consistency under different boundary conditions \cite{ShahBolechPRB16,BolechShahPRB16,FilipponeBrouwerPRB16} are avoided.}, 
the three parts of the Hamiltonian introduced above are
\begin{subequations}
\begin{eqnarray}
H_0 & = & \frac{1}{2} \int_{-\infty}^\infty dx \, 
\Big[(\partial_x {\theta})^2 + (\partial_x {\phi})^2 \Big], 
\label{eq:H0_strong1}\\[3pt]
\HT & = & \sum_\pm A_\pm \cos\big[2\sqrt{\pi}\theta(\pm\ell/2) \pm k_F\ell\big],
\label{eq:HT_boson1}\\
H_\text{gate}  & = &  \varepsilon(\Vg) \left( \frac{k_F \ell}{\pi} + \frac{1}{\sqrt{\pi}} 
\big[\theta(\ell/2) + \theta(-\ell/2) \big] \right).\quad
\label{eq:Hgate_boson1}
\end{eqnarray}
\label{eq:Hboson1}
\end{subequations}\vspace*{-0.15cm}\\
Note that the $2k_F$ backscattering in $H_\text{T}$ takes the form $\cos(2\sqrt{\pi}\theta)$ in the bosonic picture. In fact, the fermionic backscattering Eq.\,\eqref{eq:HTfermion} yields forward scattering of the bosons in addition. This can be treated exactly by redefining the field $\theta(x)$ 
\footnote{See p.\ 308 in Ref.\,\cite{GiamarchiBook}.}, 
which we assume is already done. 
However, $H_\text{gate}$ in \eqref{eq:Hgate_boson1} itself corresponds to forward scattering caused by a potential. Thus, it can be absorbed by a transformation of the fields as well; in particular, consider adding to $\theta(x)$ a linear ramp between the barriers:
\begin{equation}
\begin{aligned}
\tilde\theta(x) \equiv &\; \theta(x)  + x\, \sdfrac{\varepsilon}{\sqrt{\pi}} \Theta\left(x+\ell/2\right)\Theta\left(-x+\ell/2\right) \\
&- \sdfrac{\ell}{2}\sdfrac{\varepsilon}{\sqrt{\pi}} \Theta\left(-x-\ell/2\right) 
+ \sdfrac{\ell}{2}\sdfrac{\varepsilon}{\sqrt{\pi}}\Theta\left(x-\ell/2\right) 
\end{aligned}
\end{equation}
where $\Theta(x)$ is a step function at the origin. 
 $H_\text{gate}$ is thus absorbed into the $(\partial_x\tilde\theta)^2$ term in $H_0$, yielding
\begin{subequations}
\begin{align}
H_0 & = \frac{1}{2} \int_{-\infty}^\infty dx \, 
\Big[(\partial_x {\theta})^2 + (\partial_x {\phi})^2 \Big]  
\label{eq:H0_strong2}\\[3pt]
\HT & = \sum_\pm A_\pm 
\cos\big[2\sqrt{\pi}\theta(\pm\ell/2) \pm k_F\ell \mp \varepsilon\ell \big] 
\label{eq:HT_boson2}
\end{align}
\label{eq:Hboson2}
\end{subequations}\vspace*{-0.40cm}\\
where we have dropped the tilde on $\theta$ for clarity 
\footnote{That $\varepsilon\ell$ is dimensionless can be seen by restoring factors set to one: $\varepsilon\ell \!\to\! \varepsilon\ell/\hbar v_F$}. 


It is convenient to form the sum and difference fields, often referred to as ``charge'' and ``flavor''. These are defined on a semi-infinite 1D line, $x>0$, by
\begin{equation}
\theta_{c/f}(x) \equiv \big[ \theta(x) \,\pm\, \theta(-x) \big]/2  .
\label{eq:1d_cf_fields}
\end{equation}
The two main advantages to using $\theta_c$ and $\theta_f$ are that, first, their fluctuations are uncorrelated \cite{FendleyPRB95}, and, second, $\HT$ assumes a particularly convenient form. We find 
\begin{subequations}
\begin{align}
 H_0  & = \frac{1}{2} \!\int_0^\infty \!\!\!\!\!dx 
\Big\{ \!(\partial_x {\theta_f})^2 \!+\! (\partial_x {\phi_c})^2 +  (\partial_x {\theta_c})^2 \!+\! (\partial_x {\phi_f})^2
\Big\} , \quad\;
\label{eq:H0_strong3}\\
\HT  &= A \cos\!\left[2\sqrt{\pi}\theta_c(0)\right] \cos\!\left[ 2\sqrt{\pi}\theta_f(\ell/2) + \xi \right]
\label{eq:HT_boson3a}\\
& \; + B \sin\!\left[2\sqrt{\pi}\theta_c(0)\right] \sin\!\left[ 2\sqrt{\pi}\theta_f(\ell/2) + \xi \right],
\label{eq:HT_boson3b}\\
&\qquad\qquad\quad \text{with}\; \xi=\xi(k_F,\ell,\varepsilon)=k_F\ell - \varepsilon\ell ,
\label{eq:def-xi} 
\end{align}
\label{eq:Hboson3}
\end{subequations}\vspace*{-0.3cm}\\
where $A \!\propto\! (A_+ + A_-)$ while $B \!\propto\! (A_+ - A_-)$ is the asymmetry between the two barriers. (Because $\theta_c$ is slowly varying and nearly constant near $x\!=\!0$, its argument is shifted from $\ell/2$ to zero; on the other hand, this cannot be done for $\theta_f$ because $\theta_f(0)\!=\!0$ by definition.) 

The applied bias consists of a potential drop between particles coming from the source reservoir and those coming from the drain. The dissipative environment introduces potential fluctuations on top of this bias. The incorporation of both of these effects into the effective Hamiltonian near full transmission (strong coupling) is explained in detail in Ref.\,\cite{ZhangNoneqPRR21}. 
Here we briefly recall some important points in the derivation and then proceed by using the result. Near the full transmission fixed point, the potential drops from right-moving to left-moving particles (assuming right-moving particles come from the source as in Ref.\,\cite{ZhangNoneqPRR21}). 
Writing this in terms of fermions and then converting to the bosonic fields, one finds that the potential bias and fluctuations couple to $\theta_c$. The applied bias can be incorporated into the barrier term $\HT$ using a time-dependent gauge transformation. Upon integrating out the quadratic environment, the field $\theta_c$ becomes \emph{interacting}. Indeed, one finds LL interactions with the same relation between $g$ and $r$ as in the initial weak-link regime, namely $g=1/(1+r)$. The upshot is that the effective Hamiltonian near the full-transmission (strong-coupling) fixed point is 
\begin{subequations}
\begin{align}
\!\!\!\!H^{\text{eff}}  = &\; \sdfrac{1}{2} \int_0^{\infty} dx\, \Big\{ (\partial_x \theta_f)^2 + (\partial_x \phi_c)^2 
\label{eq:H0_strong4a}\\
 &\qquad\quad +\, (1 + r)(\partial_x \theta_c')^2 + \frac{1}{1+ r} (\partial_x \phi_f')^2\Big\} 
\label{eq:H0_strong4b}\\[3pt]
 +& A \cos\!\left[2\sqrt{\pi}\theta_c'(0) + eVt\right] \cos\!\left[ 2\sqrt{\pi}\theta_f(\ell/2) + \xi \right]
\label{eq:HT_boson4a}\\[3pt]
+& B \sin\!\left[2\sqrt{\pi}\theta_c'(0) + eVt\right] \sin\!\left[ 2\sqrt{\pi}\theta_f(\ell/2) + \xi \right] .
\label{eq:HT_boson4b}
\end{align}
\label{eq:Hboson_full}
\end{subequations}\vspace*{-0.2cm}\\
As a reminder, the bias-induced phase-shift $eVt$ in Eq.\,\eqref{eq:Hboson_full} becomes dissipation independent after the inclusion of the resistance at the wire-reservoir boundaries (see, e.g., Refs.\,\cite{PonomarenkoRPB95,SafiSchulzPRB95,MaslovStonePRB95,FrohlichPRB96,HuKaneX16} in the context of LL's).

We emphasize that the modes represented by fields $\theta_f$ and $\phi_c$ are \textit{free} while those represented by $\theta_c'$ and $\phi_f'$ are \textit{interacting}. 
The coupling between these two sets of modes is given by the barrier terms, \eqref{eq:HT_boson4a}-\eqref{eq:HT_boson4b}.
Recalling that a bosonic operator of the form $\cos (2\sqrt{\pi}\theta)$ corresponds to backscattering of the underlying fermions, we see that this coupling involves the \emph{simultaneous} backscattering of both modes. 

The resonance condition for a single-level dot is $\xi(k_F,\ell,\varepsilon) \!=\! \pi/2$ (or $k_F\ell \!=\! \pi/2 $ if $\varepsilon\!=\!0$), implying from the last term in \eqref{eq:density} that, as expected, the average number of electrons in the dot is $1/2$. 
Since we consider tuning very close to resonance, we have $|\xi  \!-\! \pi/2| \ll 1$.  

Renormalization of the system under these conditions proceeds in \emph{two stages} (see Fig.\,\ref{fig:flow}). This is best understood by finding
the temporal correlations of the barrier terms \eqref{eq:HT_boson4a}-\eqref{eq:HT_boson4b} in the interaction picture, taking \eqref{eq:H0_strong4a}-\eqref{eq:H0_strong4b} as $H_0$. Because the correlations of $\theta_c'$ and $\theta_f$ factorize and 
$\langle \sin[2\sqrt{\pi} \theta_c'(0,\tau)] \cos[2\sqrt{\pi} \theta_c'(0,0)] \rangle \!=\! 0$, there is no cross correlation between the operators in \eqref{eq:HT_boson4a} and \eqref{eq:HT_boson4b}. 
To find the correlations of $\theta_f(\ell/2)$, we first use Eq.\,\eqref{eq:1d_cf_fields} to express the operators in terms of the noninteracting 1D field $\theta(x)$ and then evaluate using standard methods 
\footnote{See pp.\ 318-323 in Ref.\,\cite{GiamarchiBook}.}. The result for large imaginary time $\tau$ is 
\begin{widetext}
\vspace*{-0.5cm}
\begin{subequations}
\begin{align}
A^2 \left\langle \cos\!\left[2\sqrt{\pi} \theta_f(\ell/2,\tau) + \xi\right] \cos\!\left[2\sqrt{\pi} \theta_f(\ell/2,0) + \xi \right]   \right\rangle 
 & = c_1 A^2 \left[ (1 + \cos 2\xi) - \frac{1}{\tau^2}  (1 - \cos 2\xi) \frac{\ell^2}{2} \right],
 \label{eq:ffield_corr1}\\
 B^2 \left\langle \sin\!\left[2\sqrt{\pi} \theta_f(\ell/2,\tau) + \xi\right] \sin\!\left[2\sqrt{\pi} \theta_f(\ell/2,0) + \xi \right]   \right\rangle 
 & = c_1 B^2 \left[ (1 - \cos 2\xi) - \frac{1}{\tau^2}  (1 + \cos 2\xi) \frac{\ell^2}{2} \right],
\label{eq:ffield_corr2}
\end{align}
\label{eq:ffield-corr}
\end{subequations}\vspace*{-0.5cm}\\
\end{widetext}
where $c_1$ is a cut-off dependent constant (here, $c_1 =  2\pi\sqrt{\ell^2 + a_0^2}/a_0$). 

The first stage of renormalization is flow toward the full-transmission (strong-coupling) fixed point. If the system is tuned to be exactly symmetric and on resonance, then $B\!=\!0$ and $\xi\!=\!\pi/2$. In this case, only the $1/\tau^2$ term in \eqref{eq:ffield_corr1} is nonzero, implying that the fluctuations of $\theta_f$ extend down to the lowest energy scales with scaling dimension one. When combined with the independent fluctuations of $\theta_c'$ [which have scaling dimension $1/(1+r)$ from the interactions in \eqref{eq:H0_strong4b}], the scaling dimension is $1 + 1/(1+r) \!>\! 1$. Hence the barrier terms are \emph{irrelevant} and the magnitude of $A$ decreases under RG. The RG flow is directly into the full-transmission fixed point \cite{ImpurityEntropy}, 
which is the QCP of this system, and there is no second renormalization stage. The QCP is therefore characterized by the bosonic Hamiltonian Eqs.\,\eqref{eq:H0_strong4a}-\eqref{eq:H0_strong4b}; because it is quadratic and spatially uniform, the transmission is unity. This case is discussed in detail in Ref.\,\cite{ZhangNoneqPRR21} where the nonequilibrium current-voltage relation is calculated at the QCP and shown to be in excellent agreement with experiment. 

In our case the system is tuned very near but \emph{not} exactly to the QCP: either $B \!\neq\!0$ or $\xi\!\neq\!\pi/2$. The correlation function of $\theta_f$ drops as  $1/\tau^2$ initially, but it eventually saturates at a constant value. Low energy fluctuations (long time) therefore do not occur---\emph{the field $\theta_f$ is frozen.} Though the RG flow starts out directed at the QCP because of the initial $1/\tau^2$ decay, when it reaches this low energy scale, the scaling dimension of the barrier terms comes only from the $\theta_c'$ factors, $1/(1+r) \!<\! 1$. Thus the barrier terms become RG \emph{relevant} and their strength begins to grow---both $A$ and $B$ (if nonzero). The flow turns away from the full-transmission fixed point, exits the vicinity of the QCP, and moves toward a stable large-barrier (weak-coupling) fixed point. This is the crossover.  

 In two limiting cases, the physical picture of this second stage is simple. For a symmetric system detuned from resonance ($B\!=\!0$ but $\xi\!\neq\!\pi/2$), the barriers become large and so Coulomb blockade effects set in, ensuring that the level on the quantum dot is either completely empty or full. Electrons traverse the dot through co-tunneling, which is a single coherent tunneling process \cite{GoldsteinPRB10, LiuRLdissipPRB14}. Such a ``single barrier'' process is indeed known to be RG relevant \cite{GiamarchiBook,GogolinBook}. 

For an asymmetric system tuned to resonance ($B\!\neq\!0$ but $\xi\!=\!\pi/2$), the barriers become more and more unequal as $B$ grows. The level in the dot is incorporated into one of the leads as that barrier gets smaller, but the other barrier completely dominates the conduction \cite{EggertAffleck92}. As in the other limiting case, the system becomes a ``single barrier'' problem. 

In both cases, the ground state at the fixed point is that the system is simply cut in two, leading to zero transmission. In the absence of transmission, the dissipation becomes disconnected, and we have arrived at the cut-wire fixed point where the impurity entropy vanishes.

\begin{figure*}
  \centering
      \includegraphics[width=1 \textwidth]{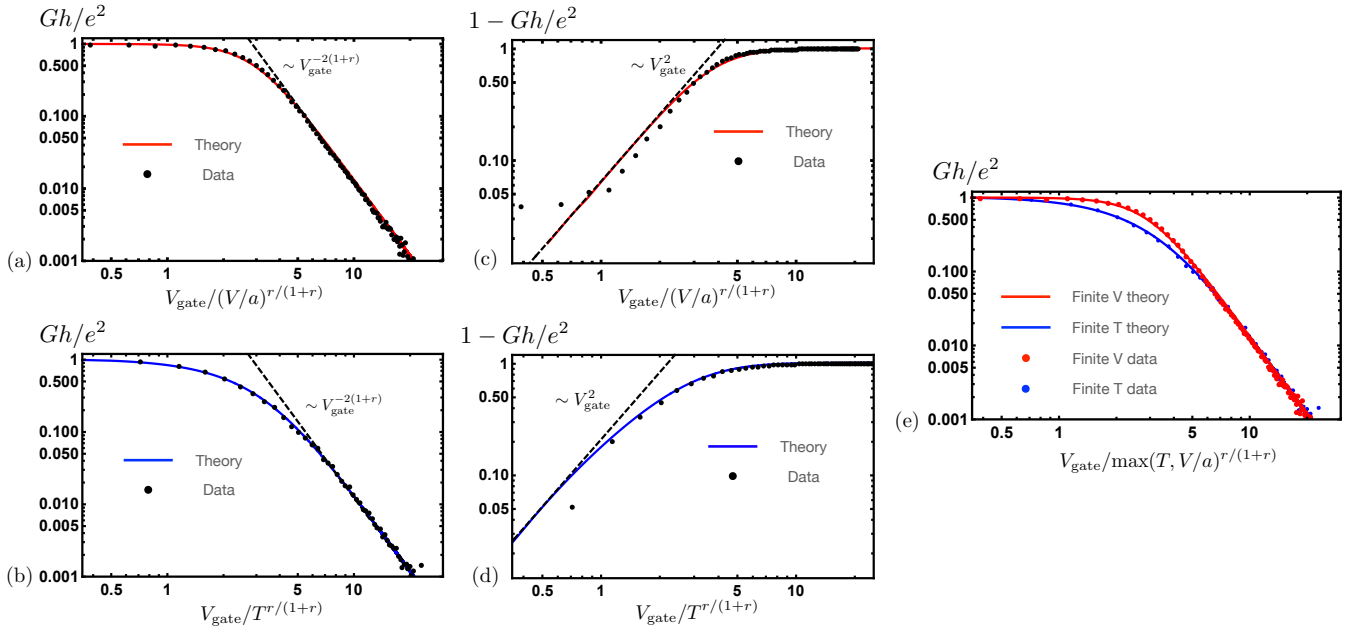}
  \caption{\label{fig:sine_gordon}
Comparison between experiment (dots) \cite{GlebData} and
theory (lines) \cite{FendleyMethod} for $r \!=\! 0.75$ [yielding an exponent of $r/(1+r)\!=\!0.43$]. 
(a)~Nonequilibrium result at zero temperature [more precisely, $T \!=\! 0.002 V$ (theory) and $T \!\approx\! 0.05 V$ (experiment)]. (b)~Equilibrium result (linear response) for finite temperature. (c),(d) Corresponding plots for the ``backscattering conductance'' $1 - Gh/e^2$.
(e) All of the data and theoretical results in a single plot. 
The constant $a\!=\! \pi\Gamma(1 + r)^{1/r} \!\approx\! 2.8$ is the relation between the bias and temperature scales. 
For the theoretical curves, we obtain $A'/\Vg\!\approx\! 0.398$ by fitting the equilibrium data in (b), thus leaving no fit parameters for the excellent agreement in the nonequilibrium case. 
Although equivalent in the two limits, in the crossover regime the temperature and bias
curves differ.}
\end{figure*}

Because the field $\theta_f(\ell/2)$ is effectively frozen during the second state of renormalization, to describe the physics of the crossover regime which is our main interest here, we remove $\theta_f(x)$ from the problem. The effective Hamiltonian that we need to study is thereby considerably simplified:
\begin{equation}
\begin{aligned}
H^{\text{eff}} = & \int_0^{\infty} \!\!dx\Big[ (1 + r)(\partial_x \theta_c')^2 + \frac{1}{1+ r} (\partial_x \phi_f')^2\Big] \\[0.2cm]
& + A' \cos\!\left[2\sqrt{\pi} \theta_c'(0) + eVt\right].
\end{aligned} 
\label{eq:bndSineGord}
\end{equation}
The effect of $\theta_f(\ell/2)$ is folded into a new coefficient of backscattering, $A'$, which undergoes the second stage of renormalization. Eq.\,\eqref{eq:bndSineGord} is the celebrated ``boundary sine-Gordon'' model \cite{GiamarchiBook,GogolinBook}. 

We close this section by summarizing our main point. Our system is accurately described by a two-stage RG process, as shown schematically in Fig.\,\ref{fig:flow}. The first stage maps the initial microscopic Hamiltonian 
[Eq.\,\eqref{eq:heff_tunneling}] to the vicinity of the full-transmission strong-coupling fixed point described by the Hamiltonian \eqref{eq:Hboson_full}. The second stage corresponds to a crossover from nearly full to very small transmission. The physical properties for a system in this crossover regime can be found by using an effective single-barrier tunneling model, namely the boundary sine-Gordon Hamiltonian, Eq.\,\eqref{eq:bndSineGord}.

\section{Nonequilibrium Current and Conductance}
\label{sec:I-V}

The nonequilibrium $I$-$V$ curve for the boundary sine-Gordon model has been investigated using a wide variety of techniques,
including the thermodynamic Bethe ansatz (TBA) \cite{FendleySaleurWarnerNPB94,FendleyPRL95,FendleyPRB95,FendleyUnifiedFrameJSP96}, fermionic methods \cite{AristovWoelflePRB14}, the nonequilibrium RG technique \cite{AristovWoelfle09}, expansion in the Coulomb gas representation (instantons) \cite{WeissBook}, and boundary conformal field theory methods \cite{GhoshalZam94,BazhanovNoneqSinGNPB99,Lukyanov_2007}. 
In this paper, we first use the TBA technique of Ref.\,\cite{FendleyPRB95} to find the differential conductance curve $G(V,T)$ and compare it with experimental data \cite{Mebrahtu13}. The theoretical results are then supported by a case that is exactly solvable.

\subsection{TBA compared to experiment: $\mathbf{ r\!=\!0.75}$}

An advantage of using the TBA technique is that it can produce the $G(V,T)$ curve for all values of the backscattering parameter $A'$ over a wide range of temperature and bias.
Strictly speaking, the TBA method is applicable only if the parameter $r$ is an integer \cite{FendleyPRB95}, a point emphasized recently \cite{boulatX2019}.
However, interestingly, it was conjectured \cite{FendleySaleur96} that this technique can be applied to systems with arbitrary values of $r$.
This conjecture is now widely accepted---for support see, for instance, Refs.\,\cite{BaurWeissPRB04,ClerkBackactionPRL06,KomnikWeissAnnPhys07} 
and compare the results of Refs.\,\cite{FendleyPRB95} and \cite{AristovWoelfle09}---and we proceed on that basis. 

The nonlinear conductance from applying the TBA method to the effective Hamiltonian Eq.\,(\ref{eq:bndSineGord}) with $r\!=\!0.75$ is shown in Fig.\,\ref{fig:sine_gordon} \cite{FendleyMethod}. 
(For convenience we use units in which $e\!=\!k_B\!=\!1$.)
Theoretical results for two limiting cases are shown: 
(i)~equilibrium, $V \!\ll\! T$ (linear response, blue lines) and (ii) far from equilibrium, $T\!\ll\! V$ (
red lines). 
Both cases are compared to experimental data from H.\ T.\ Mebrahtu et al.\ Ref.\,\cite{Mebrahtu13}  
on a system carefully tuned to be symmetric, $t_S \!=\! t_D$ \cite{GlebData}.

The independent variable in Fig.\,\ref{fig:sine_gordon} is the detuning from perfect transmission, denoted $\Vg$, because it was varied experimentally at fixed temperature and bias: the ``lineshape'' of the conductance peak was measured.  From the following scaling argument, it is natural to divide $\Vg$ by $T^{r/(1+r)}$ or $V^{r/(1+r)}$
(for the corresponding argument in a LL see Sec.\,VII of Ref.\,\cite{KaneFisherPRB92}).  Since the scaling dimension of the operator describing the deviation from the full-transmission fixed point in Eq.\,\eqref{eq:bndSineGord} is $1/(1+r)$, one expects this term to grow at low temperature or bias according to $A' \!\propto\! (T\text{ or }V)^{-[1-1/(1+r)]}$. 
General theory of critical points then suggests that the differential conductance is a universal function of $X \!\equiv\! \Vg  / (T,V/a)^{r/(1+r)}$ throughout the crossover \cite{KaneFisherPRB92}, thus justifying the $x$-axis in Fig.\,\ref{fig:sine_gordon}.  
The proportionality constant $a$ relates
the scales of $V$ and $T$. It is not a free parameter but rather is fixed by the requirement that there is a single universal behavior in the $X\!\gg\!1$ asymptotic regime: the $T\!\to\!0$ limit of linear response coincides with the $V\!\to\!0$ limit of the $T\!\ll\!V$ nonequilibrium result.  This yields 
$a \!\equiv\! \pi\Gamma(1 + r)^{1/r}$ [where $\Gamma(1 + r)$ refers to the gamma-function] \cite{KaneFisherPRB92}. 
Indeed, it was shown in Ref.\,\cite{Mebrahtu13} that the linear-response (equilibrium) conductance does obey such a universal function. 
In the nonequilibrium case, our calculation of $G(X)$ (red lines in Fig.\,\ref{fig:sine_gordon}) is the corresponding \emph{nonequilibrium universal function}. 
The direct comparison in Fig.\,\ref{fig:sine_gordon}(e) shows that the equilibrium and nonequilibrium universal functions are \emph{not} the same. 

To highlight power-law behavior, a log-log scale is used in Fig.\,\ref{fig:sine_gordon}. 
In the first column [panels (a) and (b)], note that the conductance approaches $e^2/h$ for small detuning, or equivalently for large $V$ or $T$. 
For large $X$ (small $V$ or $T$), because the detuning is not exactly zero, the conductance becomes small. In the tail ($X\!\gg\!1$, $G\!\ll\! e^2/h$), reproducing the known power-law for a single barrier, $G \!\propto\! T^{2r}$, requires therefore 
$G \!\propto\! X^{-2(1+r)} \!=\!X^{-3.5}$. 
Both the equilibrium and nonequilibrium universal curves in Fig.\,\ref{fig:sine_gordon} satisfy this constraint (dashed lines); in fact, the constant $a$ is fixed by requiring the two curves to coincide in this regime. 

To compare to experiment, there is only one free parameter in the theory, namely the magnitude of the backscattering, $A'$ in Eq.\,(\ref{eq:bndSineGord}), which is related to the experimental detuning $\Vg$. We fix this parameter by fitting the \emph{equilibrium} conductance; the best fit yields $A'/\Vg\!\approx\! 0.398$. (Changing the value of this ratio shifts the theoretical curves along the $x$-axis.) In panel (b), we see that the fit of the equilibrium universal curve to the data is excellent. 

The comparison of the \emph{nonequilibrium} theory and experiment is then \emph{parameter free}. 
First, note that in the large $X$ regime, temperature and bias are indistinguishable: the finite bias and finite temperature data points [panels (a) and (b), respectively, or panel (e) combined] lie almost on top of each other, as they should. 
The overall message from the comparison in panel (a) is that the agreement between the experimental and theoretical nonequilibrium universal curves---in the absence of any fit parameter---is remarkable. 

In order to investigate the small deviations from full transmission near the strong-coupling fixed point, we plot $1\!-\!G$ in the second column [panels (c) and (d)]. 
Near the non-Fermi-liquid QCP ($X\!\ll\!1$ and $G\!\approx\! e^2/h$), we expect the equilibrium and non-equilibrium universal curves to show the same power-law behavior. 
Since $1\!-\!G \!\sim\! |A'|^2$ by perturbation theory, one finds $1\!-\!G \!\propto\! X^2$ near the QCP (dashed lines). While the power-law is the same, the magnitude of the two curves is not: the nonequilibrium deviation from $e^2/h$ is smaller (red). Because $A'$ is set by fitting the equilibrium curve and the constant $a$ is fixed by the physical requirement that the two curves coincide in the tail, there is no freedom in the theory to adjust this difference. 

The two theory curves generally provide a good description of the experimental data in panels (c) and (d). The main deviation is for $X\!<\!1$ in the nonequilibrium case, where the experiment shows saturation and the conductance does not quite reach $e^2/h$. (With regard to the equilibrium data, see Ref.\,\cite{Mebrahtu13} for more discussion and detail.) In fact, the approach to the QCP in the experiment is being cut-off by the finite bias: when tuned exactly to the QCP ($\Vg\!=\!0$), $1 - Gh/e^2 \propto V^{2/(1+r)}$ 
\footnote{The power law here comes from the scaling dimension of the leading \emph{irrelevant} operator at the QCP, in contrast to the leading \emph{relevant} operator appearing in Eq.\,\ref{eq:bndSineGord}. See Ref.\,\cite{ZhangNoneqPRR21} for a thorough discussion of this case.}. 
While the value of $V$ used for this data prevents a close approach to the QCP thereby causing the saturation in panel (c), in the theory we used values such that the full crossover curve could be  plotted. 

In strong contrast to the two asymptotic limits, in the crossover regime centered on $X \!\sim\! 2.5$, a significant difference between the equilibrium and nonequilibrium curves emerges: here the nonequilibrium conductance has a much sharper transition between the two asymptotic regimes [Fig.\,\ref{fig:sine_gordon}(e)]. The physical reason underlying the bias/temperature difference in the crossover is discussed in Sec.\,\ref{sec:DiffCrossover}. Remarkably, the theoretical result successfully captures the experimental conductance in this regime for both the equilibrium and nonequilibrium cases, thus supporting our mapping from the DRL model to the boundary sine-Gordon model.

\begin{figure}[t]
(a)\includegraphics[width=0.85\columnwidth]{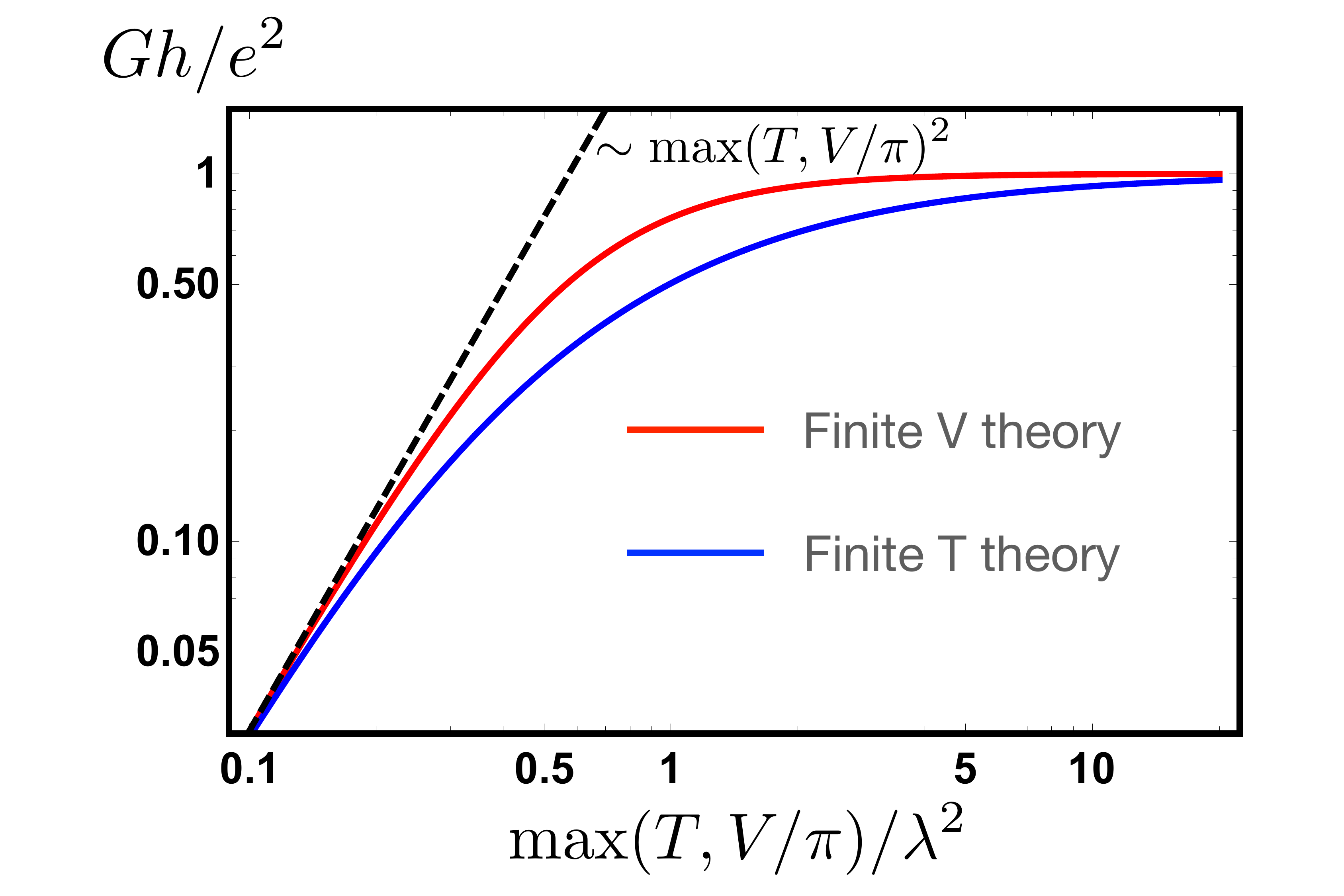}
(b)\includegraphics[width=0.85\columnwidth]{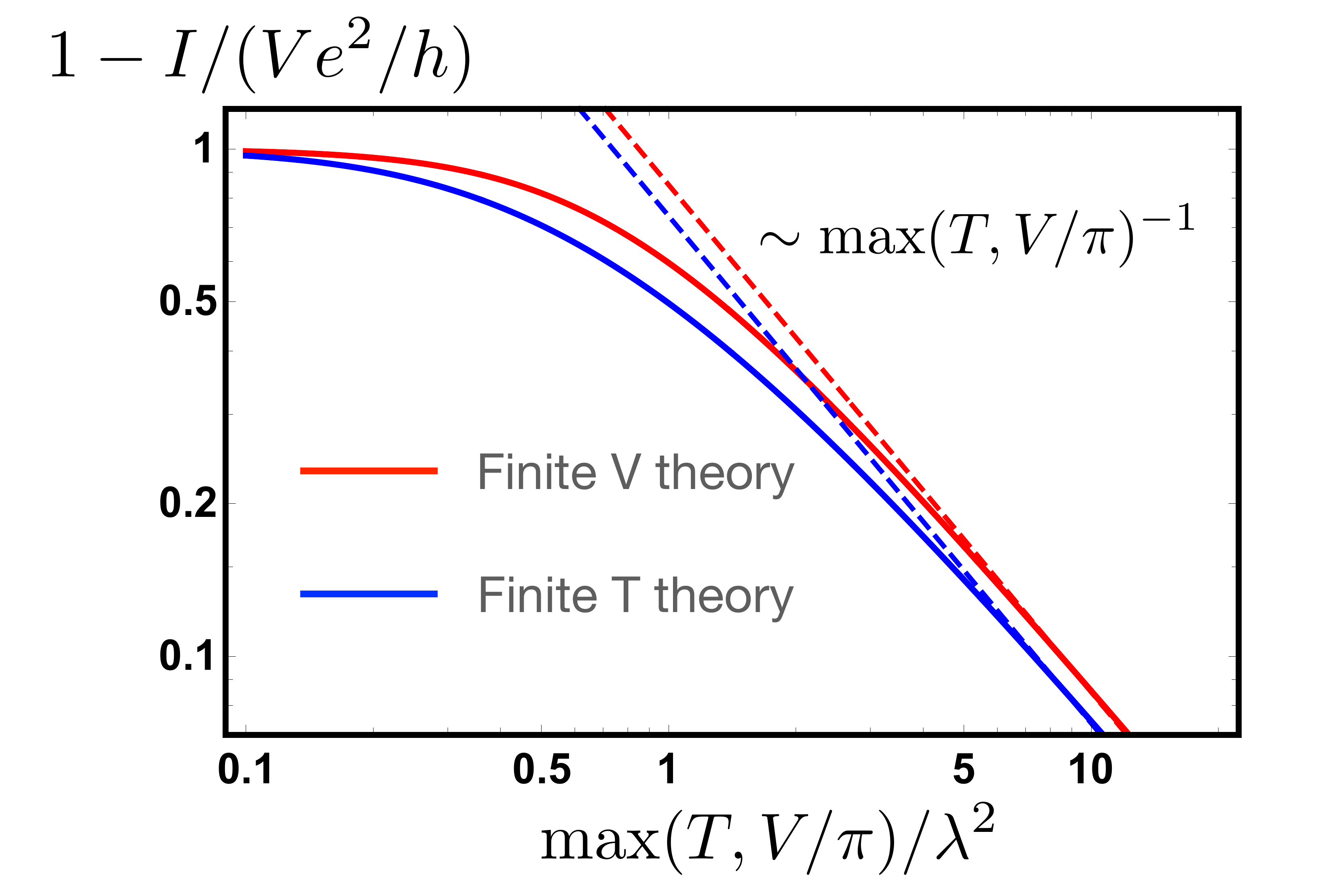}
\caption{\label{fig:toulouse_point} 
(a)~Differential conductance and (b)~``backscattering current'' for the $r\!=\!1$ case as a function of $T$ or $V$.  
The finite $V$ and finite $T$ curves correspond to the limits $V\!\gg\! T$ (where $T\!=\!10^{-7} V$) and $T\!\gg\! V$ (where $V\!=\!10^{-7} T$), respectively. We choose to present the ``backscattering current'' in (b) because the leading-order conductance is constant when the deviation is small \cite{RedAbove}.}
\end{figure}

\subsection{Exact solution: $\mathbf{r\!=\!1}$}
\label{sec:exact_expressions}

As is often the case for quantum impurity models \cite{GogolinBook}, the DRL model is exactly solvable for a specific choice of parameters.  In particular, for $r\!=\!1$ the system Hamiltonian (see App.\,\ref{sec:Hweakcoup}) can be mapped to a  Majorana resonant level model that is \emph{noninteracting}. The mapping is presented in Appendix \ref{sec:toulouse-point}; it starts with bosonization, proceeds through several unitary transformations, and finally uses refermionization. For this noninteracting problem, the nonequilibrium transport can be calculated exactly. We find in App.\,\ref{sec:toulouse-point} that the current is  
\begin{equation}
I(V,T) =  \frac{e^2}{h}\left[ V -\lambda^2
\text{Im}\,\psi \left(\frac{1}{2}+\frac{\lambda^{2}+iV}{2\pi T} \right)\right] ,
\label{eq:toulouse_current}
\end{equation}
where $\psi$ is the digamma function and $\lambda$ measures the deviation from the QCP, combining both detuning and source-drain asymmetry [defined in Eq.\,\eqref{eq:DefLambda}]. 
The nonequilibrium current \eqref{eq:toulouse_current} is similar to that in other effectively quadratic nonequilibrium mesoscopic systems, including the two-channel Kondo \cite{MitchellSelaCrossoverPRB12}, two impurity Kondo \cite{SelaExactTransPRL09}, and Toulouse-point LL \cite{KaneFisherPRB92} systems. 

In Fig.\,\ref{fig:toulouse_point} we plot the analytical $r\!=\!1$ result, Eq.\,\eqref{eq:toulouse_current}, but in a way that is slightly different from Fig.\,\ref{fig:sine_gordon}. The $x$-axis is now $1/X^2 \!\propto\! (T\text{ or }V)$, making the form of the conductance as a function of temperature or bias clearer. Thus, while Fig.\,\ref{fig:sine_gordon} highlights the lineshape as a function of detuning (the quantity measured experimentally in Ref.\,\cite{Mebrahtu13}), Fig.\,\ref{fig:toulouse_point} highlights \textit{the nonequilibrium $I$-$V$ curve}. 

This analytical result shares the main features of the $r \!=\! 0.75$ numerical result and experimental data (see App.\,\ref{sec:other_toulouse} for more plots). The exponents of the power law behavior in both asymptotic regimes are consistent with using $r\!=\!1$ in the scaling dimensions discussed above and are the same for the finite $T$ and finite $V$ curves. 
(The shift along the $x$-axis of the curves in Fig.\,\ref{fig:toulouse_point} compared to Fig.\,\ref{fig:sine_gordon} is due to matching the experimentally measured $\Vg$ in the latter, $A'/\Vg\!\approx\! 0.398$.)
Importantly, the universal nonequilibrium function is substantially different from the equilibrium function in the crossover regime: the nonequilibrium crossover is sharper.  


\section{Discussion}\label{sec:Discussion}

\subsection{Bias Crossover vs Temperature Crossover}
\label{sec:DiffCrossover}

The most interesting feature in our results is that temperature and bias lead to different behavior for the crossover (see Figs.\,\ref{fig:sine_gordon} and \ref{fig:toulouse_point}). 
Indeed, there is no reason that the equilibrium and nonequilibrium crossover functions should be the same 
\cite{KaneFisherPRB92,*KaneFisherPRB92a, AristovWoelfle09, AnthorePierrePRX18}, though at first sight one might guess from an RG perspective that these parameters are interchangeable energy cut-offs. 
Certainly, the theoretical curves agree amazingly well with the experimental data measured for the dissipation strength $r\!=\!0.75$. 

In either asymptotic regime of the conductance curves (the  weak tunneling regime where $G \!\to\! 0$ or the weak backscattering regime where $G \!\to\! e^2/h$), a single operator with a small amplitude dominates the transport. A straightforward RG argument yields power-law behavior \cite{KaneFisherPRB92}, and one simply uses $\max(T,V)$ as the RG cutoff, thereby leading to the naive notion of interchangeable energy cutoffs.  

\begin{figure}
\includegraphics[width=0.99\columnwidth]{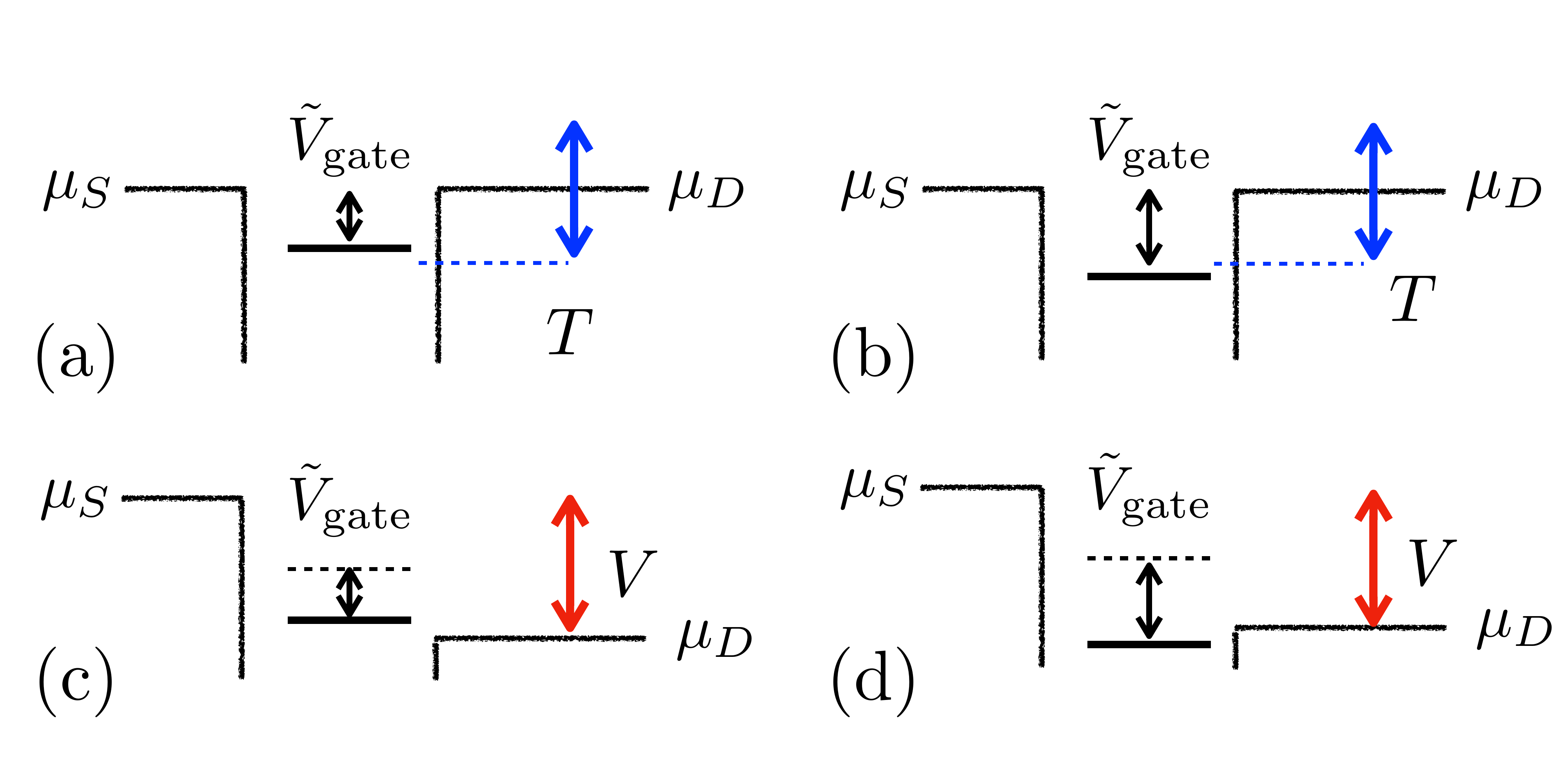}
\caption{\label{fig:illustration}
Physical picture behind the difference between temperature-induced and bias-induced crossover. The peak in the renormalized LDOS (black line) is detuned from the resonance condition by the renormalized gate voltage $\tilde{V}_\text{gate}$. The width of the energy window contributing to transport (indicated by the blue and red arrows) is associated with temperature in equilibrium (top row) or bias voltage out of equilibrium (bottom row).  
(a),(c)~The peak in the LDOS is within the energy window. (b),(d)~The LDOS peak is outside the window. The fact that the window for applied bias is a sharp rectangle makes the nonequilibrium crossover sharper. 
}
\end{figure}

The nontrivial regime is the crossover where
the perturbative RG analysis fails. 
We physically explain the temperature-bias difference in the crossover regime with the help of Fig.\,\ref{fig:illustration}.
The key ingredients are (i) how the local density of states (LDOS) in and near the quantum dot lines up with the window of energies participating in the transport and (ii) the fact that the energy window defined by temperature is smooth while that introduced by bias is a sharp rectangular window. 
Throughout the argument, it is important to keep in mind that both the detuning and the width of the resonance renormalize as $T$ or $V$ varies. We therefore consider the renormalized detuning, $\tilde{V}_\text{gate} \sim \Vg^{(1+r)/r}$, which can be compared directly to $T$ or $V$.  
First, consider the case when $\max(T,V)$ is slightly larger than the renormalized detuning, so that the weight of the LDOS is within the energy window. (In the initial system, the weight of the LDOS corresponds simply to the level in the quantum dot; near strong coupling, there is nevertheless a peak in the renormalized LDOS associated with the full-transmission condition.) In equilibrium, Fig.\,\ref{fig:illustration}(a), lead-dot tunneling is enabled only through thermal fluctuation and so is reduced compared to resonant transmission. 
In contrast, the nonequilibrium conductance, Fig.\,\ref{fig:illustration}(c), is almost unaffected by the increasing detuning. 
Compared to the finite temperature conductance, we see that when $\tilde{V}_{\text{gate}} \!<\! \max(T,V)$, the finite bias conductance is larger. 

The situation changes dramatically when the renormalized detuning becomes slightly larger than $\max(T,V)$, so that the weight of the LDOS is on the edge or in the tails of the energy window.
In the finite temperature situation Fig.\,\ref{fig:illustration}(b), there is no qualitative change---the decrease of the conductance is  similar to that in Fig.\,\ref{fig:illustration}(a).
In contrast, in the finite bias situation Fig.\,\ref{fig:illustration}(d), the weight of the renormalized LDOS is now passing outside of the sharp window, and so the conductance rapidly decreases. 
Combining this trend with that in the last paragraph, we see that in the nonequilibrium scenario, the transition in the crossover regime is much sharper.

\subsection{Connection to Two-Channel Kondo Model}\label{sec:Connection2CK}

In order to elucidate the underlying physics in quantum impurity problems, it is useful to make connections between different models. It was recognized early on that resonant tunneling in a LL gives rise to 2CK physics \cite{EggertAffleck92}. Likewise, the QCP in the DRL model is known to be two-channel Kondo-like \cite{Mebrahtu12,Mebrahtu13,ZhangNoneqPRR21}. However, an explicit mapping to the 2CK model was not made in either case. 

In Appendix \ref{sec:2CKmapping} we carry out an explicit mapping from the DRL model for $r\!=\!1$ to the 2CK model, thus showing that they are equivalent. Indeed, a spin flip in the 2CK model corresponds to a change in the occupation of the quantum dot in the DRL model, and spin current between channels in the 2CK model is equivalent to charge transfer across the dissipative dot here. The connection to the 2CK model, which is a well-known example of frustration, helps one understand why perfect symmetry is required in the DRL model to reach quantum criticality. 

The mapping is made between the bosonized forms of the two models, after a  unitary transformation. The condition  $r\!=\!1$ allows one to refermionize the bosonic model to obtain the Hamiltonian of the spin sector of the 2CK model. When  $r\!\neq\!1$, the DRL model has different, more subtle correlations than the 2CK model, ones that cannot, as far as we know, be mapped onto a noninteracting model. (See App.\,\ref{sec:toulouse-point} for mapping of the $r\!=\!1$ DRL model to a noninteracting problem.) For  $r\!\neq\!1$ the two models are qualitatively similar in terms of the underlying physics of frustration and renormalization, but are not exactly the same.

\section{Conclusion}\label{sec:Conclusion}

To summarize, in this paper we study the nonequilibrium nonlinear conductance of a dissipative resonant level as a model for a quantum dot connected to resistive leads. 
We consider a situation where slight detuning from the QCP drives the system toward the weak tunneling regime through a crossover. For the special case $r\!=\!1$ we find an exact expression for the $I(V,T)$ curve using the Keldysh formalism. 
For general $r$, we show that our problem is effectively a boundary sine-Gordon model, for which the $I(V,T)$ curve can be found through the TBA technique. Our calculation for $r \!=\! 0.75$ matches perfectly with the experimental data. 
The dependence on bias is not the same as the temperature dependence outside of the asymptotic regimes; we explain this difference by observing that the stronger thermal fluctuations in the finite temperature situation lead to a smoother transition between the two asymptotic limits.

Our argument to arrive at these results has three main steps.
First, using the RG analysis of the system Hamiltonian as a guide as well as expectations of universality connected to fixed points, we find an effective Hamiltonian, Eq.\,(\ref{eq:Hboson_full}), for the full transmission (strong coupling) regime near the QCP. 
Second, for parameters that are only slightly detuned from the QCP, an argument based on the correlation function of the resulting interaction terms \cite{GiamarchiBook} yields a simplified effective Hamiltonian for the crossover regime, Eq.\,(\ref{eq:bndSineGord}), that is simply the boundary sine-Gordon model. 
Third, we use previous work on the nonequilibrium properties of the boundary sine-Gordon model, especially results from the TBA technique by Fendley and coworkers \cite{FendleySaleurWarnerNPB94,FendleyPRB95}, to find the $I$-$V$ curve, plotted in Fig.\,\ref{fig:sine_gordon} [with the special case $r\!=\!1$ given analytically in Eq.\,(\ref{eq:toulouse_current})].
While one may question the exactness of these steps, the validity of our argument is supported by the fact that the end result is in remarkably excellent agreement with the experimental measurements. 

We close by noting that this topic ties together four themes of great interest currently:
(i)~electron correlations and non-Fermi-liquid states (though in a quantum impurity setting, not bulk), 
(ii)~nonequilibrium many-body physics in that the bias $V$ plays a distinctly different role from $T$ in the steady state, 
(iii)~open quantum systems---dissipation here causes quantum critical effects---and finally 
(iv)~quantum simulation since the dissipation generates interactions that simulate true repulsive interactions in a LL. Going forward, we believe that further developing this system---by considering, for instance, the role of spin in the quantum dot, the presence of superconductivity, or time dependence---will be even more interesting and beneficial in advancing these four themes.

\begin{acknowledgments}
We thank Gleb Finkelstein and Chung-Ting Ke  for many stimulating interactions and for providing the experimental data in Mebrahtu, et al.\ Ref.\,\cite{Mebrahtu13}. We also thank Igor Gornyi for several  helpful discussions. 
The work in the USA was supported by the U.S.\ Department of Energy, Office of Science, Office of Basic Energy Sciences, Materials Sciences and Engineering Division, under Award No.\ DE-SC0005237.
The work in Brazil was supported by FAPESP Grant 2014/26356-9. 
The work in Germany was supported by the German-Israeli Foundation for Scientific Research and Development (GIF) Grant No.\ I-1505-303.10/2019. 
\end{acknowledgments}

\appendix

\section{Initial System Hamiltonian and Bosonization\\Near the Decoupled-Level Fixed Point} 
\label{sec:Hweakcoup}

In this appendix, we first present the Hamiltonian of the system, following our previous work Ref.\,\cite{ZhangNoneqPRR21}.  This fermionic Hamiltonian is then transformed into a representation in terms of chiral bosons \cite{Mebrahtu13, LiuRLdissipPRB14}. 

The Hamiltonian for a spinless resonant level between two resistive leads consists of five parts:
\begin{equation}
H=H_\text{Dot}+H_\text{Leads}+H_\mu+H_\text{T}+H_\text{Env}.
\label{eq:H}
\end{equation}
The quantum dot is described by a single energy level which may be tuned by a gate voltage, $\epsilon_\text{d}(\Vg)$, while the source (S) and drain (D) leads contain noninteracting electrons,
\begin{equation}
H_\text{Dot}\!=\!\epsilon_\text{d}(\Vg) d^\dagger d  ,\quad
H_\text{Leads} = \sum_{\alpha=\text{S,D}} \sum_k \epsilon_k^\phdag c_{k\alpha}^\dagger c_{k\alpha}^\phdag .
\label{eq:Hdotleads}
\end{equation}

We use operators $d$ and $d^{\dagger}$ in Eq.~\eqref{eq:Hdotleads} to represent the quantum dot instead of the bosonic fields of the main text [e.g., Eq.\,\eqref{eq:Hboson1}]. This convention is taken since near the decoupled-level fixed point the dot is almost isolated and so can hardly be described by the lead fields. 
The connection between the two descriptions is made by considering a multilevel 1D quantum dot for a moment. For hard walls confining free electrons to length $\ell$, the levels would be separated by $\Delta k \!=\! \pi/\ell$ where $k$ is the wavevector in the dot. This is the same separation given by the continuum field model Eq.\,\eqref{eq:def-xi} (written in terms of the lead wavevector $k_F$), thus establishing the connection. 

Tunneling in our system excites the resistive environment through fluctuations of the voltage on the source and drain. These require a quantum description of the tunnel junction \cite{IngoldNazarov92,NazarovBlanterBook,DevoretEsteveUrbina95,VoolDevoretIJCTA17} via junction charge and phase fluctuation operators that are conjugate to each other, $\varphi_{S/D}$ and  $Q_{S/D}$. A tunneling event shifts the charge on the corresponding junction, as, for example, in this contribution to tunneling from the dot to the source:  
$c_{kS}^{\dagger}e^{-i\sqrt{2\pi}\varphi_{S}}d$. 
We take the capacitance of the two tunnel junctions to be the same and so it is natural to consider the sum and difference variables 
$\psi \equiv (\varphi_S + \varphi_D)/2$ and 
$\varphi \equiv \varphi_S - \varphi_D$.
The fluctuations $\varphi$ involve charge flow through the system and so couple to the environment. In contrast, because $\psi$ involves the total dot charge, it is not coupled to the environment \cite{IngoldNazarov92,NazarovBlanterBook,LiuRLdissipPRB14}, and we therefore drop it at this point. 
We thus arrive at the tunnel Hamiltonian
\begin{equation}
H_\text{T}= \sum_k \left( t_S^\phdag c_{kS}^{\dagger}e^{-i\sqrt{\frac{\pi}{2}} \varphi}d
+t_D^\phdag c_{kD}^{\dagger}e^{i\sqrt{\frac{\pi}{2}} \varphi}d+{\rm h.c.} \right).
\label{eq:HT}
\end{equation} 
Because of the sum over momentum, only the electrons at $x\!=\!0$ couple to the dot and environment. 
The barriers to tunneling are large in the experimental system and so  $t_S$ and $t_D$ are small---the system is in the ``weak-link'' regime near the decoupled-level fixed point (see Fig.\,\ref{fig:flow}). 

The ohmic environment of resistance $R$ is modeled in the usual way as a bath of harmonic oscillators to which the phase fluctuations $\varphi$ of the junction are coupled. The model must produce the expected temporal correlations of the phase fluctuations at long times related to the resistance of the environment by \cite{IngoldNazarov92,NazarovBlanterBook,DevoretEsteveUrbina95,VoolDevoretIJCTA17} 
\begin{equation}
\left\langle e^{-i\varphi(t)}e^{i\varphi(0)} \right\rangle \propto (1/t)^{2r} 
\quad\text{with}\quad r\equiv \frac{e^2}{h}R.
\end{equation}
One way to satisfy these constraints is to represent the environment by an infinite chiral transmission line with field $\varphi(x)$ and Hamiltonian \cite{LiuRLdissipPRB14} 
\begin{equation}
H_\text{Env} =  \frac{1}{8 r} \int_{-\infty}^{\infty} \!\!dx \,(\partial_x \varphi)^2.
\label{eq:Henv}
\end{equation}
The impedance of this transmission line is $r$. It is coupled to the junction by identifying  $\varphi(x\!=\!0)$ as the phase  $\varphi$ in the tunneling term Eq.\,\eqref{eq:HT}. 

Finally, the term driving the system out of equilibrium is  
\begin{equation}
H_\mu= \sum_{\alpha=\text{S,D}} \sum_k \mu_\alpha^\phdag c_{k\alpha}^{\dagger}c_{k\alpha}^\phdag, 
\label{eq:Hmu}
\end{equation}
where the chemical potential is related to the applied bias. Because we assume that the two junctions have the same capacitance, the voltage bias drops symmetrically, yielding $\mu_{S/D}^\phdag \!=\!\pm eV/2$. 

The renormalization of this weak-coupling model can be obtained through the method of bosonization. For the physics of the  strong-coupling fixed point discussed in Sec.\,\ref{sec:strong-coup}, it is most convenient to use bosonization in terms of canonical fields [see $\phi(x)$ and $\theta(x)$ in Eq.\,\eqref{eq:bosonization}]. In contrast, here we proceed by using chiral bosonic fields, which provides a simpler basis for presenting the exactly solvable case in App.\,\ref{sec:toulouse-point} and the mapping onto the two channel Kondo model in App.\,\ref{sec:2CKmapping}.

The source and drain leads are each represented by a semi-infinite 1D system of right- and left-moving fermions.  
(A 1D representation is possible because we are treating a quantum impurity problem, which is therefore local in space \cite{HewsonBook,GogolinBook}.) 
Each of these is then unfolded in the usual way into a right-moving field on an infinite line, 
$\psi_S (x)$ and $\psi_D (x)$, with $x\!=\!0$ at the point of coupling to the dot. We now introduce corresponding chiral bosonic fields, $\Phi_{S/D}(x)$, with the commutation relations 
 $[\partial_x \Phi_{i}(x),\,\Phi_{j}(x')] = i\delta_{ij}\pi\, \delta (x-x')$ where $i,j \in \{S,D\}$.
The fermionic and bosonic fields are related through phenomenological bosonization in the standard way \cite{GiamarchiBook,GogolinBook}, 
\begin{equation}
 \psi_{S/D}^\phdag (x) =\frac{1}{\sqrt{2\pi a_0}}F_{S/D}\exp[i\Phi_{S/D}(x)] ,
\end{equation}
where $F_{S/D}$ are the Klein factors needed to preserve the fermionic anticommutation relations, $a_0$ is a regularization scale for short distance or time, and we use the conventions of Ref.\,\cite{GiamarchiBook}. 

We now rotate the lead basis by introducing the flavor field $\Phi_{f}$ and charge field $\Phi_{c}$, defined by 
\begin{equation}
\Phi_{f} \equiv \sdfrac{1}{\sqrt{2}} (\Phi_{S}-\Phi_{D}) \quad\text{and}\quad
\Phi_{c} \equiv \sdfrac{1}{\sqrt{2}} (\Phi_{S}+\Phi_{D}).
\label{eq:CF_field}
\end{equation}
The noninteracting lead Hamiltonian is now simply
\begin{equation}
H_{\text{Leads}}=\frac{v_{F}}{4\pi}\int_{-\infty}^{\infty}dx \big[\left(\partial_{x}\Phi_{c}\right)^{2}+\big(\partial_{x}\Phi_{f}\big)^{2}\big]
\end{equation}
while the tunneling Hamiltonian (\ref{eq:HT}) becomes
\begin{equation}
\begin{aligned}
H_{\textrm{T}} &  =   t_{S}\sdfrac{F_{S}}{\sqrt{2\pi a_0}}\,
e^{-\frac{i}{\sqrt{2}}[\Phi_c+\Phi_f](x=0)} e^{-i\frac{1}{2}\varphi(0) } \,d \\ 
   & + t_{D}\sdfrac{F_{D}}{\sqrt{2\pi a_0}}\,e^{-\frac{i}{\sqrt{2}}[\Phi_c-\Phi_f](x=0)}
e^{+i\frac{1}{2}\varphi(0) } \,d
+ \text{h.c.}  
\end{aligned}
\label{eq:HT_Bosonization}
\end{equation}

A subtlety appears in treating the applied bias term: the experimentally relevant bias applied to the leads, Eq.\,\eqref{eq:Hmu}, is not the bias appropriate for the 1D wires. This has been discussed in detail in the case of a LL wire connecting reservoirs \cite{PonomarenkoRPB95, SafiSchulzPRB95, MaslovStonePRB95, FrohlichPRB96, HuKaneX16}. 
Using the same argument here [with $g\!=\!1/(1+r)$] \cite{AppC-HuKaneX16}, 
one finds that applying a 1D bias of $\pm (1+r)eV/2$ where $V$ is the physical bias yields the physical current. 
Thus, the nonequilibrium driving written using the bosonic form of the S/D density is
\begin{equation}
\begin{aligned}
H_\mu &= \frac{1+r}{2} eV \int_{-\infty}^{\infty} \!\frac{dx}{2\pi}  
\left(\partial_x \Phi_S - \partial_x \Phi_D \right) \\
& = \frac{1+r}{2\sqrt{2}\pi}eV \int_{-\infty}^{\infty} \!\!dx\, \partial_x \Phi_f.
\end{aligned}
\label{eq:bias_hamiltonian}
\end{equation}
Note that the way that $\Phi_{f}(x\!=\!0)$ and $\varphi$ appear in \eqref{eq:HT_Bosonization} is similar. Since the correlation function of both $\varphi$ and the free chiral field have a power law decay, it is natural to combine them.

The effect of the dissipative environment is incorporated into the flavor field by the transformation
\begin{align}
\Phi_{f}'(x) & \equiv \frac{1}{\sqrt{1+r}}\Big[\Phi_{f}(x)+\frac{1}{\sqrt{2}}\varphi(x) \Big],\nonumber \\
\varphi'(x) & \equiv \frac{1}{\sqrt{1+r}}\Big[\sqrt{r}\Phi_{f}(x) -\frac{1}{\sqrt{2r}}\varphi(x) \Big] .
\label{eq:absorption_f}
\end{align}
The prefactors in this transformation are chosen so that the Hamiltonian of the leads remains noninteracting and the field $\varphi'(x)$ decouples from the rest of the system. The commutation relation is preserved, 
 $[\partial_x \Phi_f'(x),\,\Phi_f'(x')] \!=\! i\pi\, \delta (x-x')$.  
 In terms of these fields, the Hamiltonian becomes
\begin{equation}
\begin{aligned}
 H &= \epsilon_\text{d}(\Vg) \,d^\dagger d 
 + \frac{\hbar v_{F}}{4\pi}\int_{-\infty}^{\infty} \!\!\!dx
 \left[\left(\partial_{x}\Phi_{c}\right)^{2}+\left(\partial_{x}\Phi_{f}'\right)^{2}\right] 
 \\
  &  +  t_S \Big[\sdfrac{F_{S}}{\sqrt{2\pi a_0}} 
 e^{-i\sqrt{\frac{1+r}{2}}\Phi_f'(0)}  e^{-i\frac{1}{\sqrt{2}}\Phi_{c}(0)} d 
 + \text{h.c.}\Big] 
\\
 &  +  t_D \Big[\sdfrac{F_{D}}{\sqrt{2\pi a_0}}  
 e^{+i\sqrt{\frac{1+r}{2}}\Phi_f'(0)}  e^{-i\frac{1}{\sqrt{2}}\Phi_{c}(0)}   d 
 + \text{h.c.}\Big] \\
 & +\sqrt{\frac{1+r}{2}} \frac{eV}{2\pi} \int_{-\infty}^{\infty} \!\!dx\, \partial_x \Phi_f', 
\end{aligned}
\label{eq:heff_tunneling}
\end{equation}
where we have dropped the decoupled term involving $\varphi'(x)$. 

Because of the way the fields have been scaled in \eqref{eq:absorption_f}, the interactions appear only in boundary terms in \eqref{eq:heff_tunneling}, namely those connected to the quantum dot. As in the strong-coupling regime discussed in Sec.\,\ref{sec:strong-coup}, \emph{one} of the fields has become interacting: $\Phi_f'$ is associated with transfer of charge through both barriers and therefore couples to the dissipative environment. This is in contrast to what happens in a LL in which both fields are interacting. 

The renormalization caused by the interactions in \eqref{eq:heff_tunneling} can be efficiently studied using the ``Coulomb gas'' RG technique developed for quantum impurity problems (see Refs.\,\cite{AndYuvalHamannPRB70,KaneFisherPRB92} for the method in general and Refs.\,\cite{SchillerIngersentEPL97,LiuRLdissipPRB14} for application to this problem). The resulting RG equations (in equilibrium) are
\begin{equation}
    \begin{aligned}
    \frac{dt_S}{d\ln \tau_c} &= t_S\left[1 - \left(\frac{1+r}{4} + K_1 + 2K_2\right)\right], \\
    \frac{dt_D}{d\ln \tau_c} &= t_D\left[1 - \left(\frac{1+r}{4} + K_1 - 2K_2\right)\right],\\
    \frac{dK_1}{d\ln\tau_c} &= -4 \tau_c^2\left[ K_1 \left(t_S^2 + t_D^2\right) + K_2 (t_S^2 - t_D^2) \right],\\
      \frac{dK_2}{d\ln\tau_c} &= -2 \tau_c^2\left[ K_2 \left(t_S^2 + t_D^2) + (t_S^2 - t_D^2\right) \right],\\
    \frac{d  \epsilon_\text{d}}{d\ln\tau_c} &= \epsilon_\text{d} 
    - 2 \left( t_S^2+ t_D^2 \right) \tau_c \sinh (\epsilon_\text{d} \tau_c),
    \end{aligned}
    \label{eq:rg_weak_ccoupling}
\end{equation}
where $K_1$ and $K_2$ are two parameters that change during the RG flow and $1/\tau_c$ is the high-energy cutoff.
The parameter $K_1$ indicates the scaling dimension of the charge field, and is thus dissipation-independent.
With its initial value $K_1 \!=\! 1$, it decreases during the RG flow, indicating the loss of the dynamics of the charge field under the influence of the charge-dot interaction.
In contrast, during the RG flow the magnitude of $K_2$ increases from its initial value of zero.
It thus increases the scaling dimension of the more weakly coupled lead, and decreases that of the stronger one. Asymmetry thus increases during the RG flow, except for symmetric coupling $t_S \!=\! t_D$ in which case $K_2$ remains zero.

In the regime $r\!<\!1$ on which we focus, the hopping to the dot in a symmetric, on-resonance system renormalizes to stronger values (corresponding to smaller tunneling barriers). This then sets the stage for the weak barrier analysis near the full-transmission (strong-coupling) quantum critical point  in Sec.\,\ref{sec:strong-coup}.

\section{Exactly Solvable Case:\\The Effectively Noninteracting (Toulouse) Point}
\label{sec:toulouse-point}

In this appendix, we treat the special case of $r \!=\! 1$, presenting an exact solution of the system Hamiltonian through refermionization of the bosonic fields. In the literature on the Kondo problem, such a point in parameter space is known as a ``Toulouse point''. For this special point, the nonequilibrium nonlinear conductance can be found analytically without resorting to the methods used in Secs.\,\ref{sec:strong-coup} and \ref{sec:I-V}.

Our starting point is the Hamiltonian Eq.\,\eqref{eq:heff_tunneling} with $r\!=\!1$. First, it is advantageous to remove the field $\Phi_c$ from the tunneling term. This is accomplished through the unitary transformation
\begin{equation}
U \equiv \exp\left[i \sdfrac{1}{\sqrt{2}} \,\Phi_c(0)\, \Big(d^{\dagger} d - \sdfrac{1}{2}\Big) \right] .
\label{eq:unitary_transformation_of_coefficients}
\end{equation}
The Hamiltonian now becomes 
\begin{align}
 H &=  \epsilon_\text{d}(\Vg) \,d^\dagger d 
 + \frac{\hbar v_{F}}{4\pi}\int_{-\infty}^{\infty} \!\!\!dx
 \left[\left(\partial_{x}\Phi_{c}\right)^{2}+\left(\partial_{x}\Phi_{f}'\right)^{2}\right]  \notag\\
& + \frac{1}{\sqrt{2\pi a_0}} \left[ t_S F_{S} e^{-i \Phi_f'(0)} d + t_D F_{D} e^{+i \Phi_f'(0)} d + \text{h.c.} \right] \notag\\
& -\frac{\hbar v_F}{2\sqrt{2}} \Big(d^{\dagger} d - \sdfrac{1}{2}\Big) \partial_x\Phi_c(0) + \frac{eV}{2\pi} \int_{-\infty}^{\infty} \!\!dx\, \partial_x \Phi_f',
\label{eq:toulouse_after_rotation}
\end{align}
after taking $r\!=\!1$.
The quartic term  produced by the unitary transformation (last line) is crucial in describing the approach to the strong-coupling fixed point \cite{Zheng1-GPRB14}. However, since it is RG irrelevant near that fixed point, the role it plays in the physics of detuning and the crossover from strong- to weak-coupling  is minor \cite{KaneFisherPRB92,GogolinBook,Zheng1-GPRB14,MitchellSelaPRL16}. It is thus dropped from further consideration \cite{note-K1quartic}. The field $\Phi_c$ is thereby decoupled from the quantum dot and so dropped as well.

A key feature of this $r \!=\! 1$ case is that the scaling dimension of the tunneling terms in Eq.\,(\ref{eq:toulouse_after_rotation}) is $1/2$, the same as that of a free fermion. Hence the entire Hamiltonian can be refermionized using $\psi_f (x)\equiv F_{S/D} e^{i\Phi_f' (x)}/\sqrt{2\pi a_0}$ \cite{GogolinBook, Zheng1-GPRB14}. 
The Hamiltonian thus becomes
\begin{equation}
\begin{aligned}
H &= \epsilon_\text{d}(\Vg) \,d^\dagger d 
 + \frac{\hbar v_{F}}{i}\int_{-\infty}^{\infty} \!\!\!dx\, \psi^\dagger_f(x) \partial_x \psi^\phdag_f(x) \\
& + \frac{t_S - t_D}{2} \big[ \psi^\dagger_f(0) - \psi^\phdag_f(0) \big] (d^{\dagger} + d) \\
& - \frac{t_S + t_D}{2} \big[ \psi^\dagger_f(0) + \psi^\phdag_f(0) \big] (d^{\dagger} - d) \\ 
&+ eV \!\int_{-\infty}^{\infty} \!\!\!dx\, \psi^\dagger_f(x)\psi^\phdag_f(x)  \;.
\end{aligned}
\label{eq:effective_tunneling_hamiltonian}
\end{equation}

We see that Majorana fermion operators occur naturally in \eqref{eq:effective_tunneling_hamiltonian} and so define them explicitly,  
\begin{subequations}
\begin{align}
a_f(x) &= \big[ \psi_f^\dagger(x) + \psi_f^\phdag(x) \big]/\sqrt{2}, \\
b_f(x) &= \big[ \psi_f^\phdag(x) - \psi_f^\dagger(x) \big]/\sqrt{2}i, \\
\chi_1^\phdag &=\big(d^{\dagger}+d\big)/\sqrt{2}, \\
\chi_2^\phdag &=\big(d-d^{\dagger}\big)/\sqrt{2}i \;.
\end{align}
\end{subequations}
The symmetric and antisymmetric combination of the hopping amplitudes are
$t_\pm \!\equiv\! t_S \!\pm\! t_D$. 
The Hamiltonian thus becomes
\begin{equation}
\begin{aligned}
H &= i\epsilon_\text{d}(\Vg)\, \chi_1^\phdag\chi_2^\phdag 
+ \frac{\hbar v_{F}}{2i} \int_{-\infty}^{\infty} \!\!\!dx\, \left( a_f \partial_x a_f + b_f \partial_x b_f \right) \\ 
&+ it_+ a_f(0)\chi_2^\phdag - it_- b_f(0)\chi_1^\phdag 
+ i eV \!\int_{-\infty}^{\infty} \!\!\!dx\, a_f(x) b_f(x)  .
\end{aligned}
\label{eq:Hmajorana}
\end{equation}
The first and fourth terms here correspond, respectively, to particle-hole and source-drain asymmetry.

The final ingredient needed is the form of the current operator in terms of the Majorana fermions. The current between the leads is 
$I \equiv i \big[(N_S - N_D)/2,\, H\big]$ where $N_{S/D}$ denote the number operators for the original fermions in the leads. After the transformation to the Majorana operators, this becomes
\begin{equation}
\hat{I}  = \frac{i}{2} t_- a_f(0)\chi_1^\phdag + \frac{i}{2} t_+ b_f (0)\chi_2^\phdag .
\label{eq:current}
\end{equation} 

The $I$-$V$ curve can be calculated from Eqs.\,\eqref{eq:Hmajorana}-\eqref{eq:current} using the standard Keldysh formalism \cite{KamenevBook}. We find
\begin{equation}
I(V,T) =  \frac{e^2}{h}\left[ V -\lambda^2
\text{Im}\,\psi \left(\frac{1}{2}+\frac{\lambda^{2}+iV}{2\pi T} \right)\right]
\label{eq:toulouse_current-app}
\end{equation}
in units of $e^{2}/h$, where $T$ is the temperature, $\psi(x)$ is the digamma function, and 
\begin{equation}
\lambda^{2} \equiv \frac{t_-^2}{2 v_F} + \frac{2v_F  \epsilon_d^2}{t_+^2} 
\label{eq:DefLambda}
\end{equation}
is the effective total deviation from the strong-coupling fixed point trajectory that combines both detuning and asymmetric leads. Eq.\,(\ref{eq:toulouse_current-app}) is used to plot the differential conductance in Fig.\,\ref{fig:toulouse_point} and is the basis for the discussion of these results at the end of Sec.\,\ref{sec:I-V}.

\section{Other Transport Curves at the Toulouse Point}
\label{sec:other_toulouse}

\begin{figure}
\includegraphics[width=0.8\columnwidth]{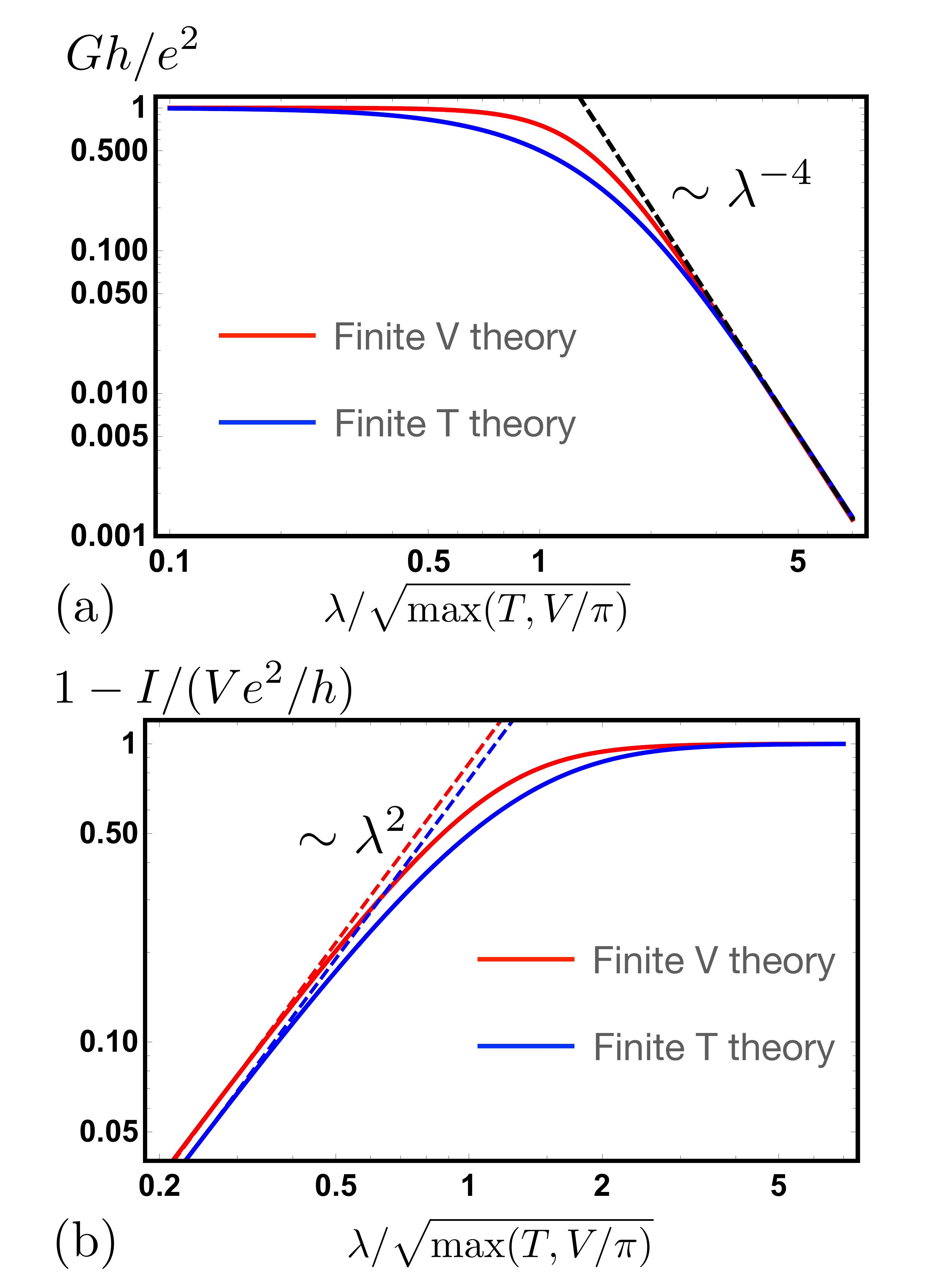}
\vspace*{-0.1cm}
\caption{\label{fig:toulouse_appendix}
(a) Conductance and (b) ``backscattering current'' as functions of the normalized detuning $\lambda/\sqrt{\max(T,V/\pi)}$, for the Toulouse point $r = 1$.
}
\end{figure}

In presenting the conductance and ``backscattering current'' for the Toulouse point $r\!=\!1$ in Sec.\,\ref{sec:exact_expressions}, we use $\max(T,V/\pi) /\lambda^2$ as the $x$-axis in Fig.\,\ref{fig:toulouse_point} rather than $X\!=\! \Vg/(T,V/a)^{r/(1+r)} \!=\!\Vg/\sqrt{(T,V/\pi)}$ as in Fig.\,\ref{fig:sine_gordon}. ($a\!=\!\pi$ for $r\!=\!1$.) As a simple alternative to the scan of $\Vg$ at fixed $T$ and $V$ used for taking the data in Fig.\,\ref{fig:sine_gordon}, this shows how the data would look if $\Vg$ were fixed and data taken as a function of temperature or bias voltage. The latter is particularly interesting---the differential conductance as a function of bias---since it is the derivative of the $I$-$V$ curve. 

However, to directly compare the $r\!=\!1$ and $r\!=\!0.75$ results, it is useful to plot the Toulouse point results in the same way as in Fig.\,\ref{fig:sine_gordon}. This is done in Fig.\,\ref{fig:toulouse_appendix}. The main features of the universal curves are clearly the same in the two cases, though of course the powerlaw seen in the asymptotic regimes depends on the value of $r$. (The crossover is centered at $X\!=\!1$ here since in the absence of experimental input $A'/\Vg\!=\!1$.) Note in particular that in the nonequilibrium curve the transition from one asymptotic regime to the other is sharper.

\section{Mapping to the Two-Channel-Kondo Model}\label{sec:2CKmapping}

In this appendix we present the mapping between the $r \!=\! 1$ DRL model and the 2CK model.
We begin with the effective Hamiltonian \eqref{eq:heff_tunneling} evaluated at $r \!=\! 1$. First, apply the unitary transformation
\begin{equation}
    U'=\exp\!\left[i  \Big(\sdfrac{1}{\sqrt{2}} - 1\Big) \, \Phi_c(0) \, 
    \Big(d^{\dagger} d - \sdfrac{1}{2}\Big) \right].
 \label{eq:rotation_map_to_2ck}
\end{equation}
This transformation 
is not the same as  \eqref{eq:unitary_transformation_of_coefficients} used at the exactly solvable point: here, we change the coefficient in the exponent from $1/\sqrt{2}$ to $1$ rather than removing it from the tunneling terms entirely. The Hamiltonian thus becomes 
\begin{align}
H =&\, H_{\text{leads}} + H_{\text{dot}} 
+ \frac{v_F}{2} \Big(1-\sdfrac{1}{\sqrt{2}}\Big) \Big(d^{\dagger} d - \sdfrac{1}{2}\Big) \partial_x \Phi_c(0) 
\notag\\
& + \lambda\Big[\sdfrac{F_{S}}{\sqrt{2\pi a_0}} e^{-i  \Phi_f'(0) - i\Phi_c(0)} d + \text{h.c.} \Big] 
\label{eq:2ck_afterunitary}\\
& + \lambda\Big[\sdfrac{F_{D}}{\sqrt{2\pi a_0}} e^{  i \Phi_f'(0) - i \Phi_c(0)} d + \text{h.c.} \Big] , \notag
\end{align}
where $\lambda$ defined in \eqref{eq:DefLambda} combines the effects of energy detuning and barrier asymmetry. In this form, the degrees of freedom represented by \emph{both} $\Phi_f'(0)$ and $\Phi_c(0)$ may be refermionized (see below).
Notice that the scaling dimension of all three lines of \eqref{eq:2ck_afterunitary} is one, as in tree-level scaling in Kondo models.  

The next step 
is to combine the quantum dot operator and Klein factors into a spin operator, 
\begin{equation}
S_- \equiv Fd, \quad S_+ \equiv d^\dagger F, \quad S_z \equiv d^\dagger d - \sdfrac{1}{2}.
\label{eq:effective_spin}
\end{equation}
In addition, the lead and environment degrees of freedom encoded in $\Phi'_f$ and $\Phi_c$ are mapped to the spin degrees of freedom of the two channels in the 2CK model. We introduce the standard 2CK labels ``s" for the total spin of both channels and ``sf" (spin flavor) for the spin imbalance between the channels (see Ref.\,\cite{SchillerHershToulousePRB98}, for instance) and identify
\begin{equation}
\Phi_c \to \Phi_{s}, \quad \Phi_f' \to \Phi_{sf}. 
\end{equation}
In this spin notation, the Hamiltonian is 
\begin{align}
H =&\, \frac{\hbar v_{F}}{4\pi}\int_{-\infty}^{\infty} \!\!\!dx
 \left[\left(\partial_{x}\Phi_{s}\right)^{2}+\left(\partial_{x}\Phi_{sf}\right)^{2}\right] \notag \\
& + BS_z + \frac{v_F}{2} \Big(1-\sdfrac{1}{\sqrt{2}}\Big)\, S_z \, \partial_x \Phi_c(0) 
\notag\\
& + \frac{\lambda}{\sqrt{2\pi a_0}}\Big[ e^{-i  \Phi_{sf}(0) - i\Phi_s(0)} S_- +  \text{h.c.} \Big] 
\label{eq:map_2ck_after}\\
& + \frac{\lambda}{\sqrt{2\pi a_0}}\Big[ e^{  i \Phi_{sf}(0) - i \Phi_s(0)} S_- + \text{h.c.} \Big] . \notag
\end{align}
This is the bosonized form of the Hamiltonian of the spin sector of the 2CK model \cite{GogolinBook,SchillerHershToulousePRB98}.

The spin-flip terms can be refermionized: the coefficient of both bosonic fields in the exponent is simply one due to the unitary transformation \eqref{eq:rotation_map_to_2ck}. The refermionization relations are \cite{GogolinBook,SchillerHershToulousePRB98}
\begin{equation}
\begin{aligned}
   & \psi^{\dagger}_{1\uparrow} \psi^\phdag_{1\downarrow} = \frac{1}{2\pi a_0} \exp[-i\ ( \Phi_s + \Phi_{sf} ) ],\\
   & \psi^{\dagger}_{2\uparrow} \psi^\phdag_{2\downarrow} = \frac{1}{2\pi a_0} \exp[-i ( \Phi_s - \Phi_{sf} ) ],\\
   & \psi^{\dagger}_{1\uparrow}\psi^\phdag_{1\uparrow} - \psi^{\dagger}_{1\downarrow}\psi^\phdag_{1\downarrow} + \psi^{\dagger}_{2\uparrow}\psi^\phdag_{2\uparrow} - \psi^{\dagger}_{2\downarrow}\psi^\phdag_{2\downarrow} = \frac{1}{\pi} \partial_x \Phi_s,
\end{aligned}
\end{equation}
where ``1" and ``2" denote the two channels of the 2CK. Eq.~\eqref{eq:map_2ck_after} then becomes
\begin{equation}
\begin{aligned}
H &= H_\text{channels} + BS_z \\
& \!\!+\! \frac{v_F  \pi}{2} \Big( 1 \!-\!\sdfrac{1}{\sqrt{2}}\Big)   S_z 
\left[\psi^{\dagger}_{1\uparrow}\psi^\phdag_{1\uparrow} 
\!-\! \psi^{\dagger}_{1\downarrow}\psi^\phdag_{1\downarrow} 
\!+\! \psi^{\dagger}_{2\uparrow}\psi^\phdag_{2\uparrow} 
\!-\! \psi^{\dagger}_{2\downarrow}\psi^\phdag_{2\downarrow} \right] \\
& \!\!+\!  \lambda\sqrt{2\pi a_0}  \left[ \psi^{\dagger}_{1\uparrow} \psi^\phdag_{1\downarrow} S_-   
+  \psi^{\dagger}_{2\uparrow} \psi^\phdag_{2\downarrow} S_-  + \text{h.c.}\right] ,
\end{aligned}
\label{eq:effective_2ck}
\end{equation}
where $H_\text{channels}$ includes not only the continuum of spin degrees of freedom of the two channels---the first line in \eqref{eq:map_2ck_after}--- but also their charge sector (which is not coupled to the local spin). 
Eq.~\eqref{eq:effective_2ck} is the Hamiltonian of the anisotropic two-channel-Kondo model. 

Importantly, it is known that the 2CK fixed point contains an impurity Majorana from quantum frustration (see, e.g., Ref.\,\cite{SchillerHershToulousePRB98}). This Majorana induces a non-trivial impurity entropy that has been recently verified through simple conductance measurement \cite{PierreMeirSelaX21}. Its interplay with a topological Majorana is also investigated in recent studies \cite{GuHaroldPRB20, GuChristianPRB20}.

\bibliography{QTransport_2021-09,BibFootnotes}

\begin{thebibliography}{132}
\bibitem [{Lar()}]{LargeR}%
  \BibitemOpen
  \href@noop {} {}\bibinfo {note} {For $1\!<\!r\!<\!3$, charge localization
  (and hence zero conductance) is possible if one starts with very weak links
  from the dot to the leads, but otherwise the physics discussed here is not
  essentially different. On the other hand, when $r\!>\!3$ all tunneling
  operators become irrelevant. This indicates that the system is stable near
  the weak-tunneling fixed point that corresponds to a doubly degenerate ground
  state---the quantum dot decouples from both leads and the charge in the dot
  is localized. See Ref.\,\cite{LiuRLdissipPRB14} for further discussion of
  these cases (in equilibrium).}\BibitemShut {Stop}%
\bibitem [{Asy()}]{AsymmetricFP}%
  \BibitemOpen
  \href@noop {} {}\bibinfo {note} {At this fixed point, the quantum dot is
  completely hybridized with the stronger lead, and the impurity entropy thus
  vanishes. In strong contrast, at the decoupled-level fixed point defined by
  $t_S \!=\! t_D \!=\! 0$, the quantum dot becomes isolated, leading to the
  impurity entropy $\ln 2$.}\BibitemShut {Stop}%
\bibitem [{Imp()}]{ImpurityEntropy}%
  \BibitemOpen
  \href@noop {} {}\bibinfo {note} {At the QCP, certain degrees of freedom
  associated with the dot become decoupled and contribute a boundary entropy.
  The best known example is the $\ln(\sqrt{2})$ entropy at the Emery-Kivelson
  point of the two-channel Kondo model due to the decoupled Majorana fermion.
  In our case, the boundary entropy can be calculated from conformal field
  theory to be $\ln(\sqrt{1+r})$
  \cite{WongAffleck94,Zheng1-GPRB14}.}\BibitemShut {Stop}%
\bibitem [{Fen()}]{FendleyMethod}%
  \BibitemOpen
  \href@noop {} {}\bibinfo {note} {More precisely, we calculate the current
  with Eqs.~(5.2) and (5.3) of Ref.\,\cite{FendleyPRB95}, after solving the
  quasiparticle spectrum following Eq.\,(4.7) of the same paper.}\BibitemShut
  {Stop}%
\bibitem [{Red()}]{RedAbove}%
  \BibitemOpen
  \href@noop {} {}\bibinfo {note} {In the lower panel, if we were to plot the
  conductance rather than the current, the blue curve would be above the red
  one.}\BibitemShut {Stop}%
\bibitem [{App()}]{AppC-HuKaneX16}%
  \BibitemOpen
  \href@noop {} {}\bibinfo {note} {Note in particular App.\,C of
  Ref.\,\cite{HuKaneX16}. Basically, the effective bias is obtained by using
  the charge conservation and no-backscattering requirements.}\BibitemShut
  {Stop}%
\bibitem [{not()}]{note-K1quartic}%
  \BibitemOpen
  \href@noop {} {}\bibinfo {note} {In addition, for the initial system near the
  decoupled-level fixed point, the decrease of the parameter $K_1$ under RG,
  Eq.\,\eqref{eq:rg_weak_ccoupling}, shows that the role of $\Phi_c$ decreases
  and so the quartic term becomes less important in the first part of the RG
  flow as well.}\BibitemShut {Stop}%
\end{thebibliography}

\end{document}